\RequirePackage[right]{lineno}
\documentclass[aps,prl,preprint,nopacs,superscriptaddress]{revtex4}
\usepackage{amsmath}
\usepackage{amssymb}
\usepackage{graphicx}
\usepackage{hyperref}
\usepackage[utf8]{inputenc}
\newcommand{\beq}{\begin{equation*}}
\newcommand{\eeq}{\end{equation*}}
\newcommand{\s}{Co$_2$MnGa}
\newcommand{\eb}{$E_\textrm{B}$}
\newcommand{\ai}{\textit{ab initio}}
\newcommand{\Ai}{\textit{Ab initio}}
\newcommand{\invA}{$\textrm{\AA}^{-1}$}

\newcommand{\pana}{A}
\newcommand{\panb}{B}
\newcommand{\panc}{C}
\newcommand{\pand}{D}
\newcommand{\pane}{E}
\newcommand{\panf}{F}
\newcommand{\pang}{G}
\newcommand{\panh}{H}
\newcommand{\pani}{I}
\newcommand{\panj}{J}
\newcommand{\pank}{K}
\newcommand{\panl}{L}
\linenumbers

\begin{document}

\title{Discovery of topological Weyl fermion lines and drumhead surface states in a room temperature magnet}

\author{Ilya Belopolski\footnote{These authors contributed equally to this work.}} \email{ilyab@princeton.edu}
\affiliation{Laboratory for Topological Quantum Matter and Spectroscopy (B7), Department of Physics, Princeton University, Princeton, New Jersey 08544, USA}

\author{Kaustuv Manna$^*$}
\affiliation{Max Planck Institute for Chemical Physics of Solids, N\"othnitzer Stra{\ss}e 40, 01187 Dresden, Germany}

\author{Daniel S. Sanchez$^*$}
\affiliation{Laboratory for Topological Quantum Matter and Spectroscopy (B7), Department of Physics, Princeton University, Princeton, New Jersey 08544, USA}

\author{Guoqing Chang$^*$}
\affiliation{Laboratory for Topological Quantum Matter and Spectroscopy (B7), Department of Physics, Princeton University, Princeton, New Jersey 08544, USA}

\author{Benedikt Ernst}
\affiliation{Max Planck Institute for Chemical Physics of Solids, N\"othnitzer Stra{\ss}e 40, 01187 Dresden, Germany}

\author{Jiaxin Yin}
\affiliation{Laboratory for Topological Quantum Matter and Spectroscopy (B7), Department of Physics, Princeton University, Princeton, New Jersey 08544, USA}

\author{Songtian S. Zhang}
\affiliation{Laboratory for Topological Quantum Matter and Spectroscopy (B7), Department of Physics, Princeton University, Princeton, New Jersey 08544, USA}

\author{Tyler A. Cochran}
\affiliation{Laboratory for Topological Quantum Matter and Spectroscopy (B7), Department of Physics, Princeton University, Princeton, New Jersey 08544, USA}

\author{Nana Shumiya}
\affiliation{Laboratory for Topological Quantum Matter and Spectroscopy (B7), Department of Physics, Princeton University, Princeton, New Jersey 08544, USA}

\author{Hao Zheng}
\affiliation{Laboratory for Topological Quantum Matter and Spectroscopy (B7), Department of Physics, Princeton University, Princeton, New Jersey 08544, USA}

\author{Bahadur Singh}
\affiliation{SZU-NUS Collaborative Center and International Collaborative Laboratory of 2D Materials for Optoelectronic Science $\&$ Technology, College of Optoelectronic Engineering, Shenzhen University, Shenzhen 518060, China}

\author{Guang Bian}
\affiliation{Department of Physics \& Astronomy, University of Missouri, Columbia, Missouri 65211, USA}

\author{Daniel Multer}
\affiliation{Laboratory for Topological Quantum Matter and Spectroscopy (B7), Department of Physics, Princeton University, Princeton, New Jersey 08544, USA}

\author{Maksim Litskevich}
\affiliation{Laboratory for Topological Quantum Matter and Spectroscopy (B7), Department of Physics, Princeton University, Princeton, New Jersey 08544, USA}

\author{Xiaoting Zhou}
\affiliation{Department of Physics, National Cheng Kung University, Tainan 701, Taiwan}

\author{Shin-Ming Huang}
\affiliation{Department of Physics, National Sun Yat-Sen University, Kaohsiung 804, Taiwan}

\author{Baokai Wang}
\affiliation{Department of Physics, Northeastern University, Boston, Massachusetts 02115, USA}

\author{Tay-Rong Chang}
\affiliation{Department of Physics, National Cheng Kung University, Tainan 701, Taiwan}
\affiliation{Center for Quantum Frontiers of Research \& Technology (QFort), Tainan, 701, Taiwan}

\author{Su-Yang Xu}
\affiliation{Laboratory for Topological Quantum Matter and Spectroscopy (B7), Department of Physics, Princeton University, Princeton, New Jersey 08544, USA}

\author{Arun Bansil}
\affiliation{Department of Physics, Northeastern University, Boston, Massachusetts 02115, USA}

\author{Claudia Felser}
\affiliation{Max Planck Institute for Chemical Physics of Solids, N\"othnitzer Stra{\ss}e 40, 01187 Dresden, Germany}

\author{Hsin Lin}
\affiliation{Institute of Physics, Academia Sinica, Taipei 11529, Taiwan}

\author{M. Zahid Hasan} \email{mzhasan@princeton.edu}
\affiliation{Laboratory for Topological Quantum Matter and Spectroscopy (B7), Department of Physics, Princeton University, Princeton, New Jersey 08544, USA}
\affiliation{Princeton Institute for Science and Technology of Materials, Princeton University, Princeton, New Jersey, 08544, USA}
\affiliation{Materials Sciences Division, Lawrence Berkeley National Laboratory, Berkeley, CA 94720, USA}


\begin{abstract}
Topological matter is known to exhibit unconventional surface states and anomalous transport owing to unusual bulk electronic topology. In this study, we use photoemission spectroscopy and quantum transport to elucidate the topology of the room temperature magnet \s. We observe sharp bulk Weyl fermion line dispersions indicative of nontrivial topological invariants present in the magnetic phase. On the surface of the magnet, we observe electronic wave functions that take the form of drumheads, enabling us to directly visualize the crucial components of the bulk-boundary topological correspondence. By considering the Berry curvature field associated with the observed topological Weyl fermion lines, we quantitatively account for the giant anomalous Hall response observed in our samples. Our experimental results suggest a rich interplay of strongly correlated electrons and topology in this quantum magnet.
\end{abstract}

\date{\today}
\maketitle

The discovery of topological phases of matter has led to a new paradigm in physics, which not only explores the analogs of particles relevant for high energy physics, but also offers new perspectives and pathways for the application of quantum materials \cite{news_Castelvecchi,ReviewQuantumMaterials_KeimerMoore,ReviewQuantumMaterials_Hsieh,ReviewQuantumMaterials_Nagaosa,RMPTopoBandThy_Bansil,Review_ZahidSuyangGuang,ARCMP_me,RMPWeylDirac_Armitage,PbTaSe2_Guang,Co2MnGa_Guoqing}. To date, most topological phases have been discovered in non-magnetic materials \cite{Review_ZahidSuyangGuang,RMPWeylDirac_Armitage,ARCMP_me}, which severely limits their magnetic field tunability and electronic/magnetic functionality. Identifying and understanding electronic topology in magnetic materials will not only provide indispensable information to make their existing magnetic properties more robust, but also has the potential to lead to the discovery of novel magnetic response that can be used to explore future spintronics technology. Recently, several magnets were found to exhibit a large anomalous Hall response in transport, which has been linked to a large Berry curvature in their electronic structures \cite{Co2MnGa_Kaustuv,Co2MnGa_Nakatsuji,Co3Sn2S2_Enke,Fe3GeTe2_Kim_Pohang,Fe3Sn2_Jiaxin}. However, it is largely unclear in experiment whether the Berry curvature originates from a topological band structure, such as Dirac/Weyl point or line nodes, due to the lack of spectroscopic investigation. In particular, there is no direct visualization of a topological magnetic phase demonstrating a bulk-boundary correspondence with associated anomalous transport.

Here we use angle-resolved photoemission spectroscopy (ARPES), \ai\ calculation and transport to explore the electronic topological phase of the ferromagnet \s\ \cite{Co2MnGa_Guoqing}. In our ARPES spectra we discover a line node in the bulk of the sample. Taken together with our \ai\ calculations, we conclude that we observe Weyl lines protected by crystalline mirror symmetry and requiring magnetic order. In ARPES we further observe drumhead surface states connecting the bulk Weyl lines, revealing a bulk-boundary correspondence in a magnet. Combining our ARPES and \ai\ calculation results with transport, we further find that Berry curvature concentrated by the Weyl lines accounts for the giant intrinsic anomalous Hall response in \s.

\begin{figure}
\centering
\includegraphics[width=14cm,trim={1in 4.1in 1in 1in},clip]{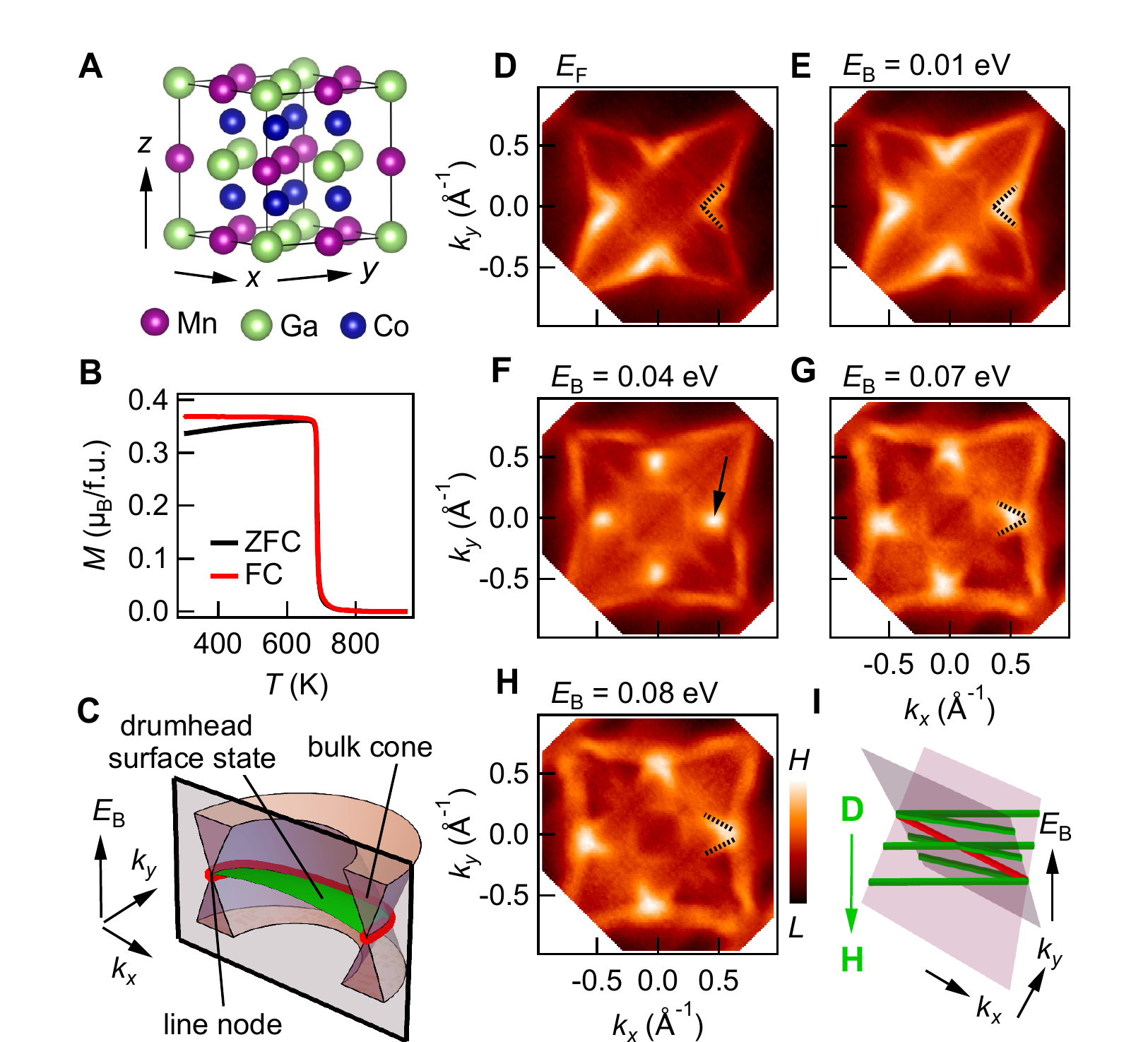}
\caption{\label{Fig1} {\bf Magnetic line node in \s.} ({\bf \pana}) Crystal structure of \s. ({\bf \panb}) Magnetization as a function of temperature of \s\ single crystals, in the absence of a magnetic field (zero-field-cooled, ZFC) and cooled under a constant magnetic field of $\mu_0 H = 200$ Oe oriented along the [001] crystallographic axis (field-cooled, FC). We find a Curie temperature $T_C = 690$ K. ({\bf \panc}) Schematic of a generic line node. A line node (red curve) is a band degeneracy along an entire curve in the bulk Brillouin zone. It is associated with a drumhead surface state stretching across the line node (green sheet). In the case of a mirror-symmetry-protected line node, the line node lives in a mirror plane of the Brillouin zone, but it is allowed to disperse in energy. ({\bf \pand-\panh}) Constant-energy surfaces of \s\ measured by ARPES at $h \nu = 50$ eV and $T = 20$ K, presented at a series of binding energies, \eb, from the Fermi level, $E_\textrm{F}$, down to \eb\ = 0.08 eV. ({\bf \pani}) Schematic of constant-energy cuts (green curves) of a line node, suggesting a correspondence with the observed ARPES dispersion.}
\end{figure}

Weyl lines can be understood within a simple framework where one categorizes a topological phase by the dimensionality of the band touching: there are topological insulators, point node semimetals and line node semimetals \cite{ClassificationGapped_Schnyder,RMPWeylDirac_Armitage,ClassificationReflection_Schnyder,TINI_Balents}. Point nodes are often further sub-categorized as Dirac points, Weyl points and other exotic point touchings \cite{RMPWeylDirac_Armitage}. Analogously, line nodes can include Dirac lines (four-fold degenerate), Weyl lines (two-fold degenerate) and possibly other one-dimensional band crossings \cite{WeylDiracLoop_Nandkishore,WeylLoopSuperconductor_Nandkishore,WeylLines_Kane}. Line nodes can be protected by crystal mirror symmetry, giving rise to drumhead surface states \cite{TINI_Balents,Ca3P2_Schnyder,ClassificationReflection_Schnyder,PbTaSe2_Guang,NodalChain_Soluyanov,RMPClassification_ChingKai,DiracLineNodes_WeiderKane}. \s\ takes the full-Heusler crystal structure (Fig. \ref{Fig1}\pana), with a cubic face-centered Bravais lattice, space group $Fm\bar{3}m$ (No. 225), indicating the presence of several mirror symmetries in the system. Moreover, the material is ferromagnetic with Co and Mn moments \cite{Co2MnGa_neutron} and Curie temperature $T_\textrm{C} = 690$ K (Fig. \ref{Fig1}\panb) \cite{Co2MnGa_CurieTemp}, indicating broken time-reversal symmetry. This suggests that all bands are generically singly-degenerate and that mirror symmetry may give rise to two-fold degenerate line nodes. In a detailed theoretical analysis, we studied the band structure of \s\ by \ai\ calculation, neglecting spin-orbit coupling (SOC). We observed that the ferromagnetic exchange splitting drives a phase with two majority spin bands near the Fermi level that exhibit two-fold degeneracies on the mirror planes [18]. These degeneracies, which arise due to a crossing of bands with opposite mirror eigenvalues, form three families of Weyl lines (Fig. \ref{Fig1}\panc, Fig. S10), which are pinned to each other, forming a nodal chain, and some of which further form Hopf-like links with one another. The predicted Weyl lines are protected only when the spin-orbit coupling (SOC) is strictly zero, but numerical results in the presence of SOC suggest that the gap opened is negligible (Fig. S9).

\begin{figure}
\centering
\includegraphics[width=16cm,trim={1in 5.3in 1in 1in},clip]{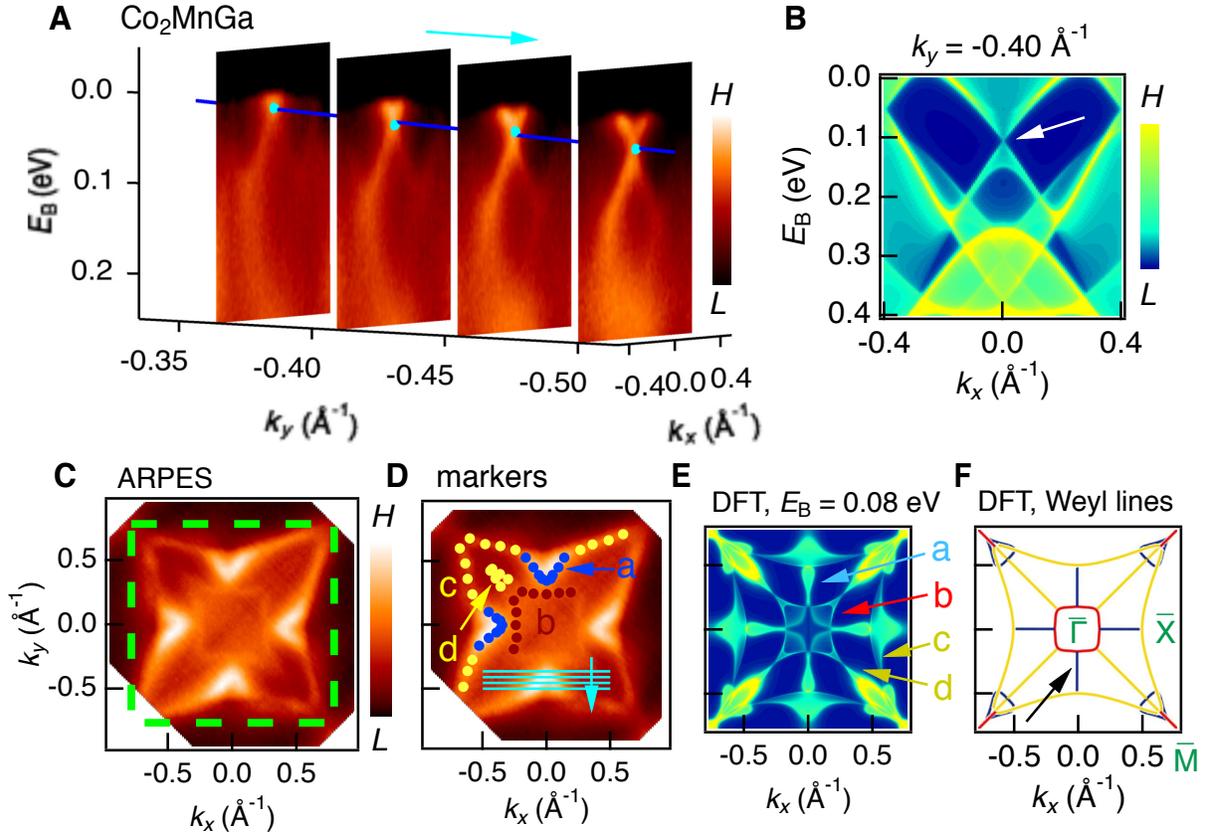}
\caption{\label{Fig2} {\bf Evidence for a Weyl line.} ({\bf \pana}) Series of ARPES $E_\textrm{B}-k_x$ cuts through the candidate line node, corresponding to the feature discussed in Fig. \ref{Fig1}. The band crossing points near $k_x = 0$ are fit with a single Lorentzian peak (cyan dots) and the train of dots is then fit with a line (blue line), the experimentally-observed line node dispersion. ({\bf \panb}) \textit{Ab initio} $E_\textrm{B}-k_x$ predicted bulk bands of \s\ in the ferromagnetic state, projected on the (001) surface, predicting a Weyl line at $k_x = 0$ (white arrow) \cite{Co2MnGa_Guoqing}. The colors indicate the spectral weight of a given bulk state on the surface, obtained using an iterative Green's function method \cite{GreensFunction_Bryant}. ({\bf \panc}) Same as Fig. \ref{Fig1}\pane, with (001) surface Brillouin zone marked (green box). ({\bf \pand}) Key features of the data, obtained from analysis of the momentum and energy distribution curves of the ARPES spectrum. ({\bf \pane}) \textit{Ab initio} constant-energy surface at binding energy $E_\textrm{B} = 0.08$ eV below $E_\textrm{F}$, on the (001) surface with MnGa termination, showing qualitative agreement with the ARPES, as marked by $a-d$. ({\bf \panf}) Projection of the predicted Weyl lines on the (001) surface, with energy axis collapsed, suggesting that the key features observed in ARPES and DFT arise from the predicted Weyl lines: $a$ (blue Weyl line), $b$ (red), $c$ (yellow), $d$ (another copy of the yellow Weyl line).}
\end{figure}

Motivated by these considerations, we investigate \s\ single crystals by ARPES. We focus first on the constant energy surfaces at different binding energies, \eb. We readily observe a feature which exhibits an unusual evolution from a $<$ shape (Fig. \ref{Fig1}\pand,\pane) to a dot (Fig. \ref{Fig1}\panf) to a $>$ shape (Fig. \ref{Fig1}\pang-\panh). This feature suggests that we observe a pair of bands which touch at a series of points in momentum space. As we shift downward in \eb, the touching point moves from left to right (black guides to the eye) and we note that at certain \eb\ (Fig. \ref{Fig1}\panf) the spectral weight appears to be dominated by the crossing point. This series of momentum-space patterns is characteristic of a line node (Fig. \ref{Fig1}\pani). For the constant-energy surfaces of a line node, as we slide down in \eb\ the touching point slides from one end of the line node to the other, gradually zipping closed an electron-like pocket (upper band) and unzipping a hole-like pocket (lower band). To better understand this result, we consider $E_\textrm{B}-k_x$ cuts passing through the line node feature (Fig. \ref{Fig2}\pana). On these cuts, we observe a candidate band crossing near $k_x = 0$. We further find that this crossing persists in a range of $k_y$ and moves downward in energy as we cut further from $\bar{\Gamma}$ (more negative $k_y$). We can fit the candidate band crossing with a single Lorentzian peak, suggestive of a series of touching points between the upper and lower bands (Fig. S13). Taking these fitted touching points, we can in turn fit the dispersion of the candidate line node to linear order, obtaining a slope $v = 0.079 \pm 0.018\ \textrm{eV\AA}$. Lastly, we observe that at a given $k_y$, the bands disperse linearly in energy away from the touching points. In this way, our ARPES results suggest the presence of a line node at the Fermi level in \s.

To better understand our experimental results, we compare our spectra with an \textit{ab initio} calculation of \s\ in the ferromagnetic state \cite{Co2MnGa_Guoqing}. We consider the spectral weight of bulk states on the (001) surface and we study an $E_\textrm{B}-k_x$ cut in the region of interest (Fig. \ref{Fig2}\panb) \cite{SM}. At $k_x = 0$ we observe a band crossing (white arrow) which we can trace back in numerics to a line node near the $X$ point of the bulk Brillouin zone (blue line node in Fig. S10). According to our earlier theoretical analysis, it arises on the $M_{xy}$ (and equivalent) mirror planes \cite{Co2MnGa_Guoqing}. This line node is a Weyl line, in the sense that it is a two-fold degenerate band crossing extended along one dimension \cite{WeylDiracLoop_Nandkishore,WeylLoopSuperconductor_Nandkishore,WeylLines_Kane}. It is predicted to be pinned to a second, distinct Weyl line, forming part of a nodal chain. To compare experiment and theory in greater detail, we plot the calculated dispersion of the Weyl line against the dispersion as extracted from Lorentzian fits of ARPES data (Fig. S13). We observe a hole-doping of experiment relative to theory of $E_\textrm{B} = 0.08 \pm 0.01 \ \textrm{eV}$. We speculate that this shift may be due to a chemical doping of the sample or an approximation in the way that DFT captures magnetism in this material. The correspondence between the crossing observed in \textit{ab initio} calculation and ARPES suggests that we have observed a magnetic Weyl line in \s.

Having considered the blue line node, we search for other line nodes in our data. We compare an ARPES spectrum (Fig. \ref{Fig2}\panc,\pand) to an \textit{ab initio} calculation of the surface spectral weight of bulk states, taking into account the observed effective hole-doping of our sample (Fig. \ref{Fig2}\pane). In addition to the blue Weyl line (labelled here as $a$), we observe a correspondence between three features in experiment and theory: $b$, $c$ and $d$. To better understand the origin of these features, we consider all of the predicted Weyl lines in \s\ \cite{Co2MnGa_Guoqing} and we plot their surface projection with the energy axis collapsed (Fig. \ref{Fig2}\panf). We observe a correspondence between $b$ in the ARPES spectrum and the red Weyl line. Similarly, we see that $c$ and $d$ match with predicted yellow Weyl lines. To further test this correspondence, we look again at our ARPES constant-energy cuts and we find that $d$ exhibits a $<$ to $>$ transition suggestive of a line node (Figs. S16, S17). The comparison between ARPES and \ai\ calculation suggests that an entire network of magnetic Weyl lines is realized in \s.

\begin{figure}
\centering
\includegraphics[width=17cm,trim={1.1in 6.6in 1in 1in},clip]{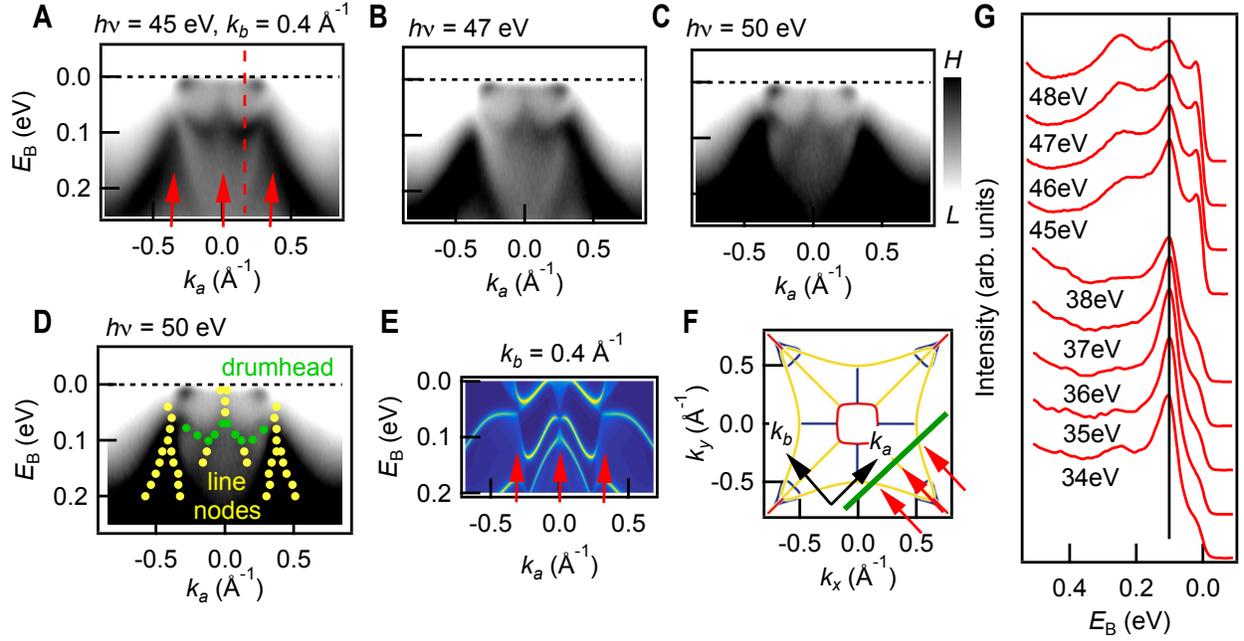}
\caption{\label{FigDrum} \textbf{Drumhead surface states in Co$_2$MnGa.} (\textbf{\pana}-\textbf{\panc}) ARPES $E_\textrm{B}-k_a$ cuts at different photon energies. We observe three cone-like features (red arrows). (\textbf{\pand}) Key features of the data of \panc, obtained from analysis of the momentum and energy distribution curves (MDCs/EDCs). Apart from the cone-like features (yellow) there are additional states (green) connecting the cones. (\textbf{\pane}) The corresponding $E_\textrm{B}-k_a$ cut from \textit{ab initio} calculation, crossing three Weyl lines (red arrows) connected by drumhead surface states \cite{SM}. (\textbf{\panf}) Same as Fig. \ref{Fig2}\panf, marking the location of the ARPES spectra in \textbf{\pana}-\textbf{\panc} (green line) and defining the $k_{a,b}$ axes. (\textbf{\pang}) Photon energy dependence of an EDC passing through the candidate drumhead state (red dotted line in \pana). The peaks marked by the black vertical line correspond to the drumhead surface state. We observe no dispersion as a function of photon energy, providing evidence that the candidate drumhead is a surface state.}
\end{figure}

Next we explore the topological surface states. We study the ARPES spectrum along $k_a$, as marked by the green line in Fig. \ref{FigDrum}\panf. On this cut we observe three cones (red arrows in Fig. \ref{FigDrum}\pana) which are consistent with the yellow Weyl lines. Interestingly, we also observe a pair of states which appear to connect one cone to the next (Fig. \ref{FigDrum}\pana-\panc). Moreover, these extra states consistently terminate on the candidate yellow Weyl lines as we vary $k_b$ (Fig. S20). We further carry out a photon energy dependence and we discover that these extra states do not disperse with photon energy from $h \nu = 34$ to $48$ eV, suggestive of a surface state (Fig. \ref{FigDrum}\pang). In \ai\ calculation, we observe a similar pattern of yellow Weyl lines pinning a surface state (Fig. \ref{FigDrum}\pane) \cite{SM}. These observations suggest that we have observed a drumhead surface state stretching across Weyl lines in \s. The pinning of the surface states to the cones further points to a bulk-boundary correspondence between the bulk Weyl lines and the drumhead surface state dispersion.

\begin{figure}
\centering
\includegraphics[width=15cm,trim={0in 3.3in 0in 0.4in},clip]{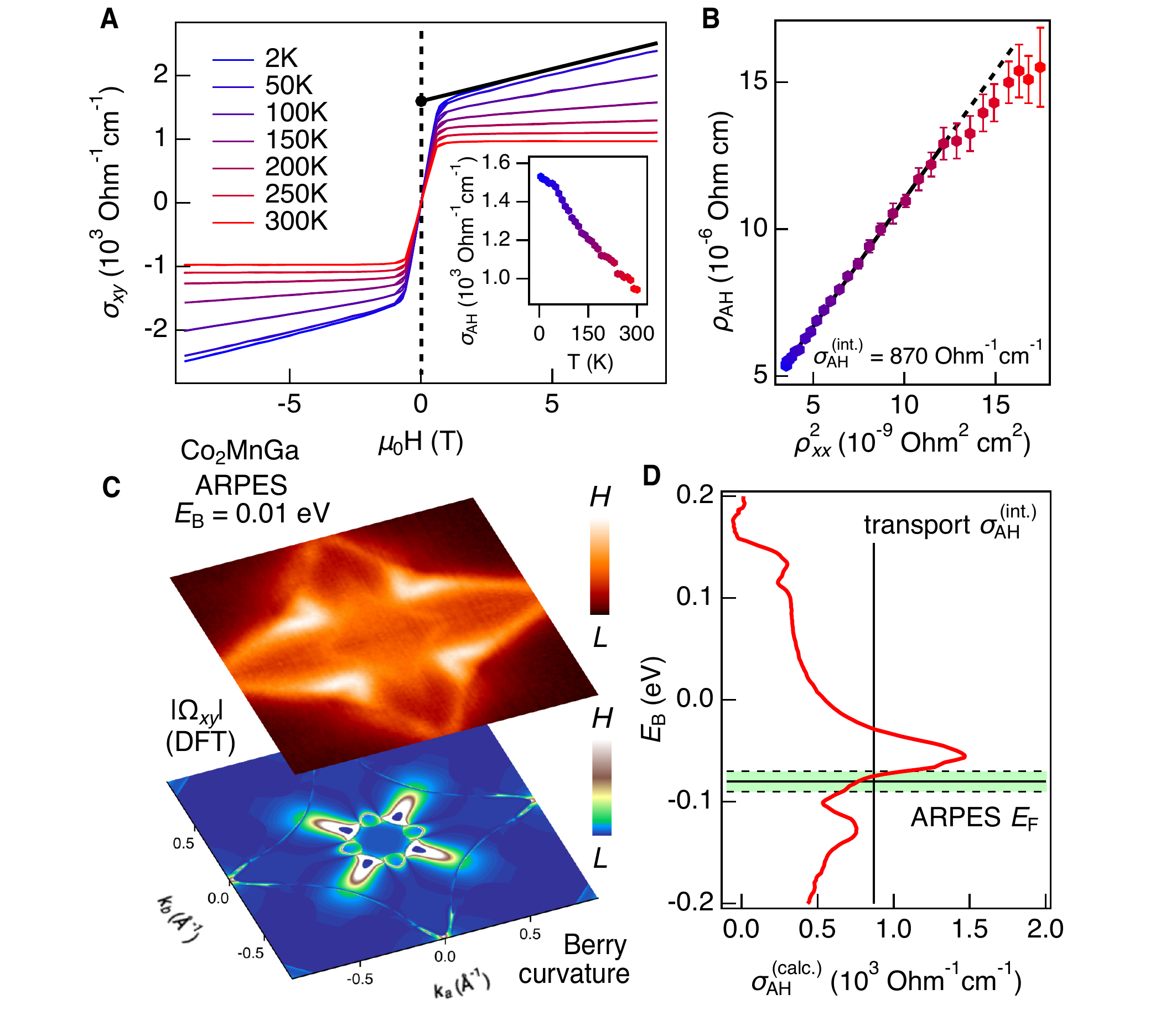}
\caption{\label{AHE} {\bf Giant anomalous Hall transport and topological Weyl lines.} ({\bf \pana}) The Hall conductivity $\sigma_{xy}$, measured as a function of applied magnetic field $\mu_0 H$ at several representative temperatures $T$, after two-point averaging of the raw data (Fig. S5), with $\mu_0 H$ applied along $[110]$ and current along $[001]$. Inset: the anomalous Hall conductivity, $\sigma_{\textrm{AH}}(T)$, obtained from $\sigma_{xy}$. ({\bf \panb}) The anomalous Hall resistivity $\rho_{\textrm{AH}}$ plotted against $\rho_{xx}^2$, both as functions of $T$, as indicated by the colors: blue (2 K) $\rightarrow$ red (300 K). A linear scaling relation estimates the intrinsic, Berry curvature contribution to the AHE, given by the slope of the line \cite{ScalingAHE_TianYeJin_2009,ScalingAHE_TianJinNiu_2015,ScalingAHE_TianJin_2016}. ({\bf \panc}) Bottom: $z$-component of the Berry curvature, calculated with spin-orbit coupling, integrated up to $E_\textrm{B} = -0.09$ eV, $|\Omega_{xy}|$. Top: the ARPES constant-energy surface at the corresponding \eb\ (same as Fig. \ref{Fig1}\pand). The correspondence between ARPES and DFT suggests that the Berry curvature is dominated by the Weyl lines. ({\bf \pand}) 
Prediction of $\sigma^{\textrm{int}}_{\textrm{AH}}$ by integrating the Berry curvature from DFT up to a given \eb\ (red curve), with $E_\textrm{F}$ set from ARPES and compared to the estimated $\sigma^{\textrm{int}}_{\textrm{AH}}$ from transport. The ARPES/DFT prediction is consistent with transport, suggesting that the line nodes dominate the giant, intrinsic AHE in \s.}
\end{figure}

Now that we have provided spectroscopic evidence for a magnetic bulk-boundary correspondence in \s, we investigate the relationship between the topological line nodes and the anomalous Hall effect (AHE). We study the Hall conductivity $\sigma_{xy}$ under magnetic field $\mu_0 H$ at different temperatures $T$ and we extract the anomalous Hall conductivity $\sigma_\textrm{AH} (T)$ (Fig. \ref{AHE}\pana). We obtain a very large AHE value of $\sigma_{\textrm{AH}} = 1530 \ \Omega^{-1} \ \textrm{cm}^{-1}$ at 2 K, consistent with earlier reports \cite{Co2MnGa_Kaustuv,Co2MnGa_Nakatsuji}. To understand the origin of the large AHE, we study the scaling relation between the anomalous Hall resistivity, $\rho_\textrm{AH}$, and the square of the longitudinal resistivity, $\rho_{xx}^2$, both considered as a function of temperature. It has been shown that under the appropriate conditions, the scaling relation takes the form,
\beq
\rho_\textrm{AH} = (\alpha \rho_{xx0} + \beta \rho^2_{xx0}) + \gamma \rho^2_{xx},
\eeq
where $\rho_{xx0}$ is the residual longitudinal resistivity, $\alpha$ represents the contribution from skew scattering, $\beta$ represents the side-jump term and $\gamma$ represents the intrinsic Berry curvature contribution to the AHE \cite{ScalingAHE_TianYeJin_2009,ScalingAHE_TianJinNiu_2015,ScalingAHE_TianJin_2016}. When we plot $\rho_\textrm{AH}$ against $\rho_{xx}^2$, we observe that a linear scaling appears to hold below $\sim 230$ K (Fig. \ref{AHE}\panb). It is possible that the deviation from linearity at high temperature arises from cancellations of Berry curvature associated with thermal broadening of the Fermi-Dirac distribution, as recently proposed for the AHE in metals \cite{ScalingAHE_YeTianJinXiao_2012}. From the linear fit, we find that the intrinsic Berry curvature contribution to the AHE is $\gamma = 870 \ \Omega^{-1} \ \textrm{cm}^{-1}$. This large intrinsic AHE leads us to consider the role of the Weyl lines in producing a large Berry curvature. To explore this question, we compare the intrinsic AHE measured in transport with a prediction based on ARPES and DFT. We observe in first-principles that the Berry curvature distribution, which we calculate in the presence of spin-orbit coupling, is dominated by the topological line nodes (Fig. \ref{AHE}\panc) \cite{SM}. Next we integrate the Berry curvature up to a given binding energy to predict $\sigma^\textrm{int}_{\textrm{AH}}$ as a function of the Fermi level. Then we set the Fermi level from ARPES, predicting $\sigma^\textrm{int}_{\textrm{AH}} = 770_{- 100}^{+ 130} \ \Omega^{-1} \ \textrm{cm}^{-1}$ (Fig. \ref{AHE}\pand). This is in remarkable agreement with the value extracted from transport, suggesting that the topological line nodes contribute significantly to the large AHE in \s.

In summary, our ARPES and corresponding transport experiments, supported by \ai\ calculation, provide evidence for magnetic Weyl lines in the room-temperature ferromagnet \s. We further find that the Weyl lines give rise to drumhead surface states and a large anomalous Hall response, providing the first demonstration of a topological magnetic bulk-boundary correspondence with associated anomalous transport. Since there are 1651 magnetic space groups and thousands of magnets in three-dimensional solids, the experimental methodology of transport-bulk-boundary exploration established here can be a valuable guideline in probing and discovering novel topological phenomena on the surfaces and the bulk of magnetic materials.

\section{Acknowledgments}

{\bf Funding:} Work at Princeton University is supported by the U.S. Department of Energy (DOE) under Basic Energy Sciences, grant no. DOE/BES DE-FG-02-05ER46200. I.B. acknowledges the support of the Harold W. Dodds Fellowship of Princeton University. The work at Northeastern University was supported by DOE, Office of Science, Basic Energy Sciences grant no. DE-FG02-07ER46352 and benefited from Northeastern University's Advanced Scientific Computation Center (ASCC) and the NERSC supercomputing center through DOE grant no. DE-AC02-05CH11231. Use of the Stanford Synchrotron Radiation Lightsource (SSRL), SLAC National Accelerator Laboratory, is supported by the U.S. Department of Energy, Office of Science, Office of Basic Energy Sciences, under contract no. DE-AC02-76SF00515. This research used resources of the Advanced Light Source, which is a DOE Office of Science User Facility under contract no. DE-AC02-05CH11231. K.M., B.E., and C.F. acknowledge the financial support by the ERC Advanced Grant no. 291472 ``Idea Heusler'' and 742068 ``TOPMAT.'' T.-R.C was supported by the Young Scholar Fellowship Program of the Ministry of Science and Technology (MOST) of Taiwan, under MOST Grant for the Columbus Program no. MOST 108-2636-M-006-002, National Cheng Kung University, Taiwan, and the National Center for Theoretical Sciences (NCTS), Taiwan. This work was further partially supported by the MOST, Taiwan, grant no. MOST 107-2627-E-006-001. This research was supported in part by the Higher Education Sprout Project, Ministry of Education to the Headquarters of University Advancement at National Cheng Kung University. STM characterization of samples was supported by the Gordon and Betty Moore Foundation (GBMF4547/Hasan). M.Z.H. acknowledges support from the Miller Institute of Basic Research in Science at the University of California at Berkeley and Lawrence Berkeley National Laboratory in the form of a Visiting Miller Professorship during the early stages of this work. M.Z.H. also acknowledges visiting scientist support from IQIM at the California Institute of Technology.\\

{\bf Scientific support:} We thank D. Lu and M. Hashimoto at Beamlines 5-2 and 5-4 of the Stanford Synchrotron Radiation Lightsource (SSRL) at the SLAC National Accelerator Laboratory, CA, USA, and J. Denlinger at Beamline 4.0.3 and S.-K. Mo at Beamline 10.0.1 of the Advanced Light Source (ALS) at Lawrence Berkeley National Laboratory (LBNL), CA, USA, for support.\\

{\bf Author contributions:} The project was initiated by I.B., K.M., C.F., and M.Z.H. I.B. and D.S.S. carried out ARPES measurements and analyzed ARPES data with assistance from G.C., J.Y., S.S.Z., T.C., N.S., H.Z., G.B., D.M., M.L., S.-Y.X., and M.Z.H. K.M. synthesized the single-crystal samples and carried out the transport and magnetization measurements with assistance from B.E. I.B. and K.M. analyzed the transport data with assistance from G.C. and T.C. G.C. performed first-principles calculations and theoretical analysis with assistance and guidance from B.S., G.B., X.Z., S.-M.H., B.W., T.-R.C., S.-Y.X., A.B., and H.L. The project was supervised by C.F., H.L., and M.Z.H. All authors discussed the results and contributed to writing the manuscript.\\

{\bf Competing interests:} The authors declare no competing interests.\\

{\bf Data and materials availability:} The data presented in this work are available on Zenodo \cite{Source}.

\section{Supplementary Materials}

References \cite{DFT2,DFT3,DFT4,Co2MnGa_exchange,Co2MnAl_Jakob,ReviewAHE_Nagaosa_Ong}.


\clearpage

\clearpage

\clearpage

\clearpage

\clearpage
\begin{center}
\textbf{\large Supplementary Materials for:\\
\ \\
Discovery of topological Weyl fermion lines and drumhead surface states in a room temperature magnet}
\end{center}

\setcounter{equation}{0}
\setcounter{figure}{0}
\setcounter{table}{0}
\makeatletter
\renewcommand{\theequation}{S\arabic{equation}}
\renewcommand{\thefigure}{S\arabic{figure}}
\renewcommand{\thetable}{S\arabic{table}}

\author{Ilya Belopolski$^*$} 

\author{Kaustuv Manna$^*$}

\author{Daniel S. Sanchez$^*$}

\author{Guoqing Chang$^*$}

\author{Benedikt Ernst}

\author{Jiaxin Yin}

\author{Songtian S. Zhang}

\author{Tyler A. Cochran}

\author{Nana Shumiya}

\author{Hao Zheng}

\author{Bahadur Singh}

\author{Guang Bian}

\author{Daniel Multer}

\author{Maksim Litskevich}

\author{Xiaoting Zhou}

\author{Shin-Ming Huang}

\author{Baokai Wang}

\author{Tay-Rong Chang}

\author{Su-Yang Xu}

\author{Arun Bansil}

\author{Claudia Felser}

\author{Hsin Lin}

\author{M. Zahid Hasan}  

\maketitle


\section{Materials and Methods}

\subsection{Single crystal growth}

\begin{figure*}
\centering
\includegraphics[width=13cm]{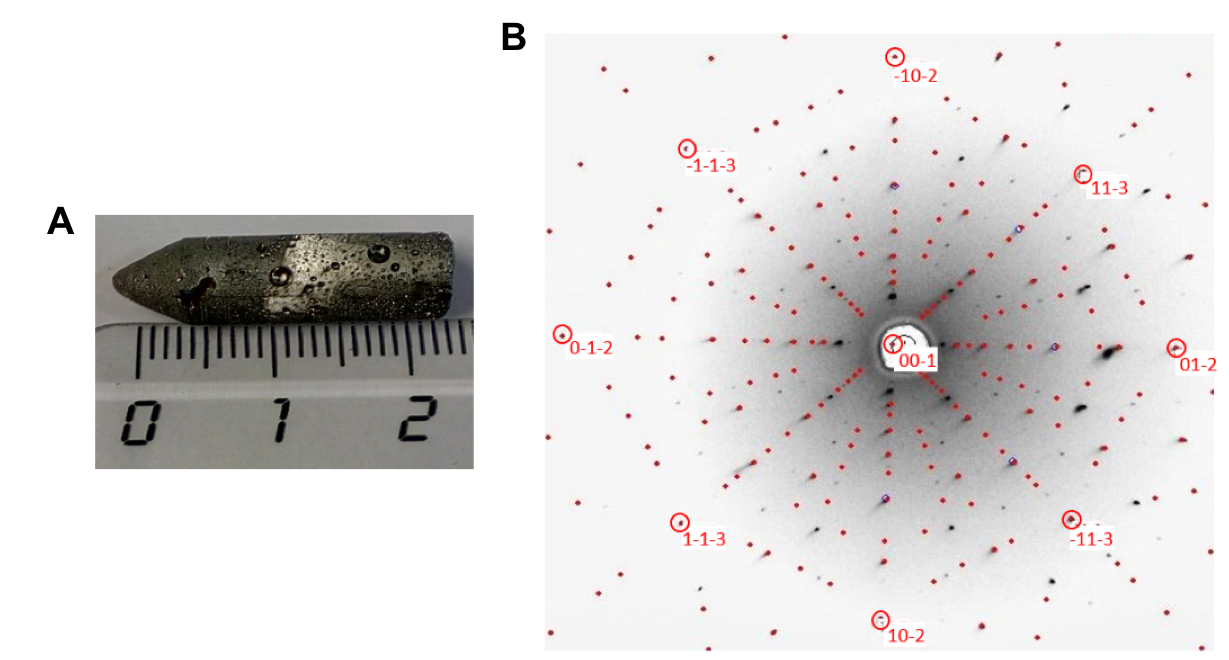}
\caption{\label{FigS1} \textbf{Crystal structure of Co$_2$MnGa.} (\textbf{\pana}) Grown single crystal of the full Heusler material Co$_2$MnGa. (\textbf{\panb}) Laue diffraction pattern of a [001] oriented crystal superposed with a theoretically simulated one.}
\end{figure*}

Single crystals of Co$_2$MnGa were grown using the Bridgman-Stockbarger crystal growth technique. First, we prepared a polycrystalline ingot using the induction melt technique with the stoichiometric mixture of Co, Mn and Ga metal pieces of 99.99\% purity. Then, we poured the powdered material into an alumina crucible and sealed it in a tantalum tube. The growth temperature was controlled with a thermocouple attached to the bottom of the crucible. For the heating cycle, the entire material was melted above 1200$^{\circ}$C and then slowly cooled below 900$^{\circ}$C (Fig. \ref{FigS1}\pana). We analyzed the crystals with white beam backscattering Laue X-ray diffraction at room temperature (Fig. \ref{FigS1}\panb). The samples show very sharp spots that can be indexed by a single pattern, suggesting excellent quality of the grown crystals without any twinning or domains. We show a representative Laue diffraction pattern of the grown Co$_2$MnGa crystal superimposed on a theoretically-simulated pattern, Fig. S1B. The crystal structure is found to be cubic $Fm\bar{3}m$ with lattice parameter $a=5.771(5)$ $\textrm{\AA}$.

\subsection{Magnetization, transport}

Magnetic measurements were performed using a Quantum Design vibrating sample magnetometer (VSM) operating in a temperature range of $2 - 950$ K with magnetic field up to 7 T. The transport experiments were performed in a Quantum Design physical property measurement system (PPMS, ACT option) in a temperature range of $2 - 350$ K with magnetic field up to 9 T. For the longitudinal and Hall resistivity measurements, we employed a 4-wire and 5-wire geometry, respectively, with a 25 $\mu$m platinum wire spot-welded on the surface of the oriented \s\ single crystals.

\subsection{Angle-resolved photoemission spectroscopy}

\begin{figure*}
\centering
\includegraphics[width=16cm,trim={1in 6.9in 1in 1in},clip]{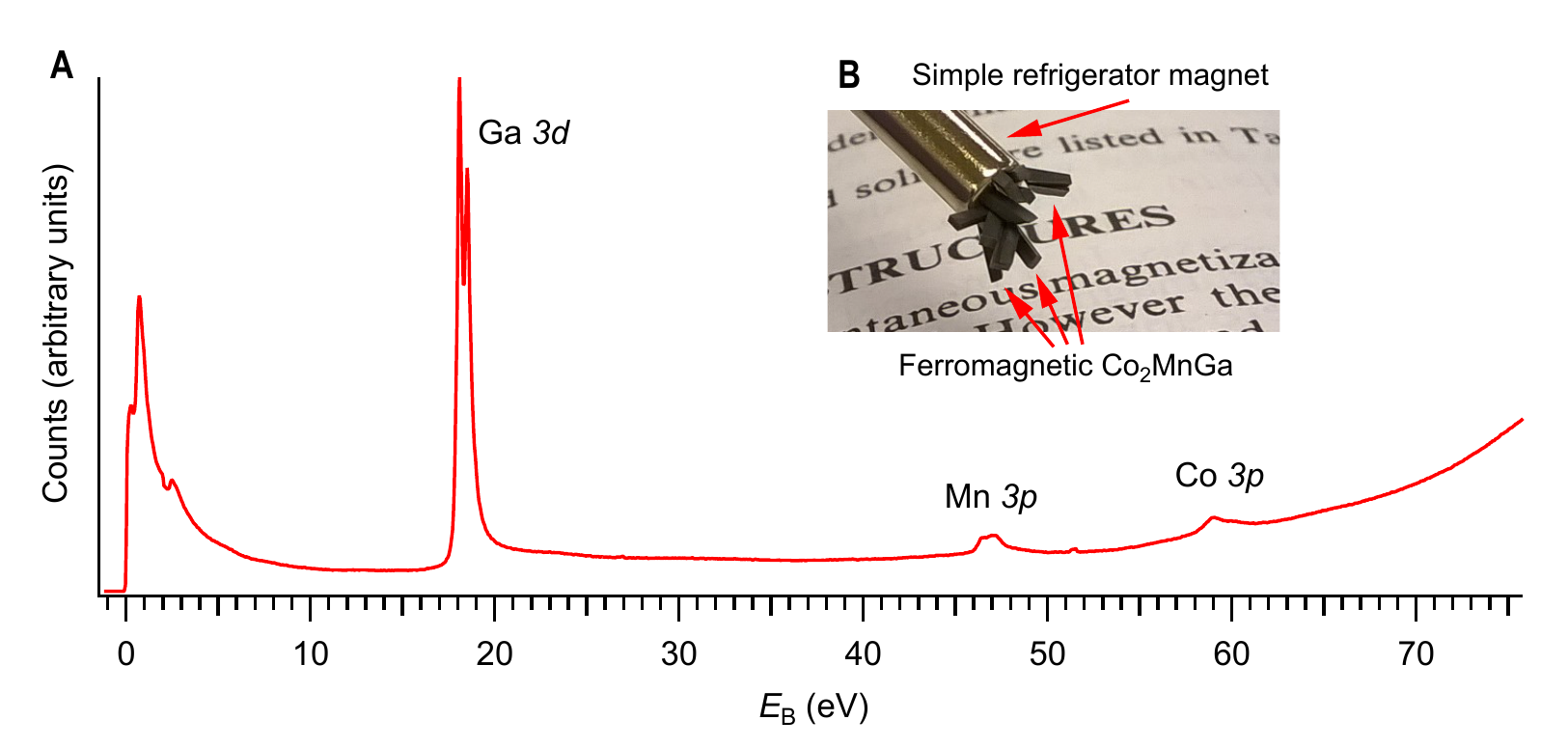}
\caption{\label{FigCore} \textbf{Core level spectrum of Co$_2$MnGa.} (\textbf{\pana}) An XPS spectrum of our Co$_2$MnGa samples clearly shows Ga $3d$, Mn $3p$ and Co $3p$ peaks, without significant irrelevant core level peaks, suggesting that our samples are of high quality. (\textbf{\panb}) The single crystal Co$_2$MnGa samples are readily picked up by an ordinary refrigerator magnet at room temperature.}
\end{figure*}

Ultraviolet ARPES measurements were carried out at Beamlines 5-2 and 5-4 of the Stanford Synchrotron Radiation Lightsource, SLAC in Menlo Park, CA, USA with a Scienta R4000 electron analyzer. The angular resolution was better than 0.2$^{\circ}$ and the energy resolution better than 20 meV, with a beam spot size of about $50 \times 40$ $\mu$m for Beamline 5-2 and $100 \times 80$ $\mu$m for Beamline 5-4. Samples were cleaved $\textit{in situ}$ and measured under vacuum better than $5 \times 10^{-11}$ Torr at temperatures $<$ 25 K. A core level spectrum of \s\ measured with 100 eV photons showed peaks consistent with the elemental composition (Fig. S\ref{FigCore}).

\subsection{First-principles calculations}

Numerical calculations of Co$_{2}$MnGa were performed within the density functional theory (DFT) framework using the projector augmented wave method  as implemented in the VASP package \cite{DFT2, DFT3}. The generalized gradient approximation (GGA) \cite{DFT4} and a $\Gamma$-centered $k$-point $12 \times 12 \times 12$ mesh were used. Ga $s, p$ orbitals and Mn, Co $d$ orbitals were used to generate a real space tight-binding model, giving the Wannier functions. The surface states on a (001) semi-infinite slab were calculated from the Wannier functions by an iterative Green's function method.

\section{Supplementary Text}

\subsection{Magnetism \& transport}

\begin{figure*}
\centering
\includegraphics[width=16cm]{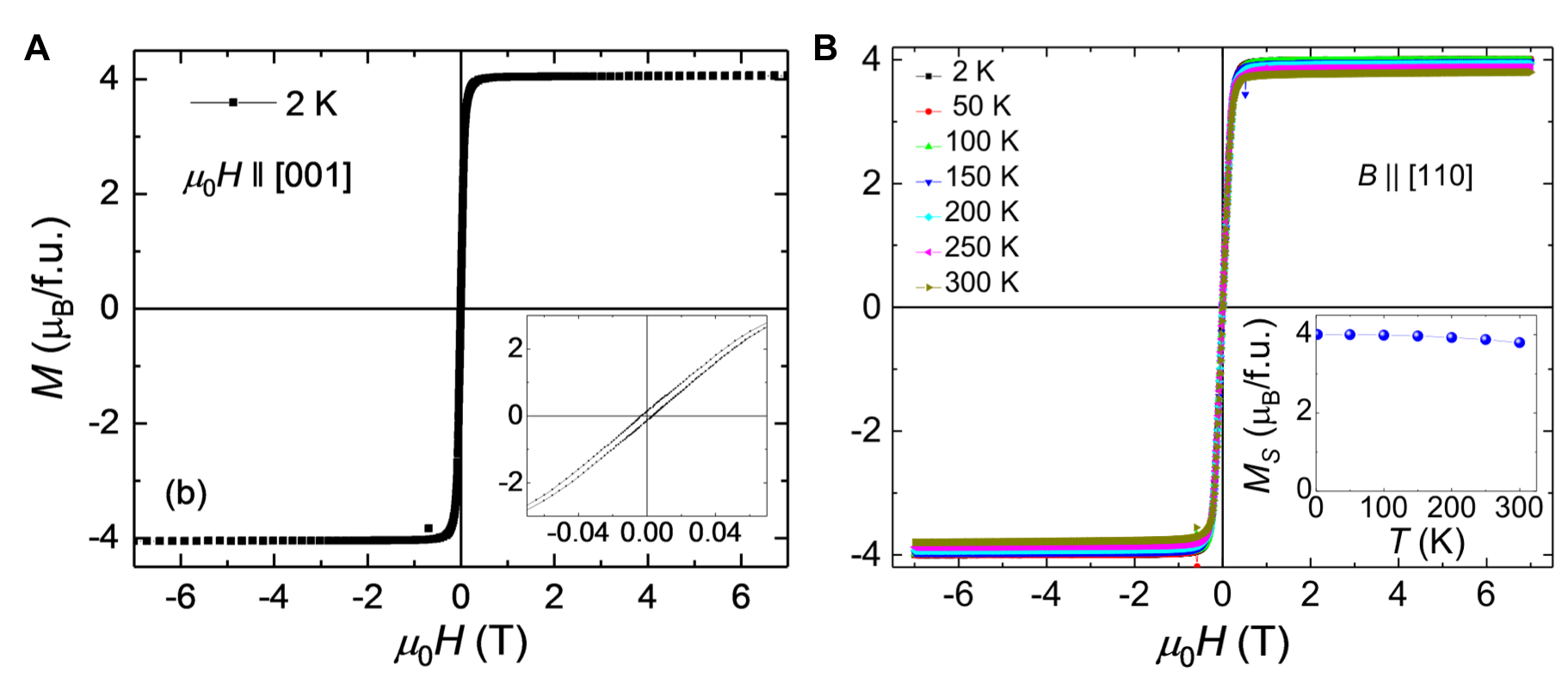}
\caption{\label{FigS2} \textbf{Magnetic hysteresis of Co$_2$MnGa.} (\textbf{\pana}) Hysteresis loop at 2 K for a [001] oriented Co$_2$MnGa crystal. Inset: zoomed-in view at low field, with hysteresis loop. (\textbf{\panb}) Hysteresis loop for various temperatures with field applied along the [110] direction. Inset: temperature dependence of the saturation magnetization, which decreases slightly from $4.0 \ \mu_B$ at 2 K to $3.8 \ \mu_B$ at 300 K.}
\end{figure*}

\begin{figure*}
\centering
\includegraphics[width=9.5cm,trim={0in 0.35in 0in 0.8in},clip]{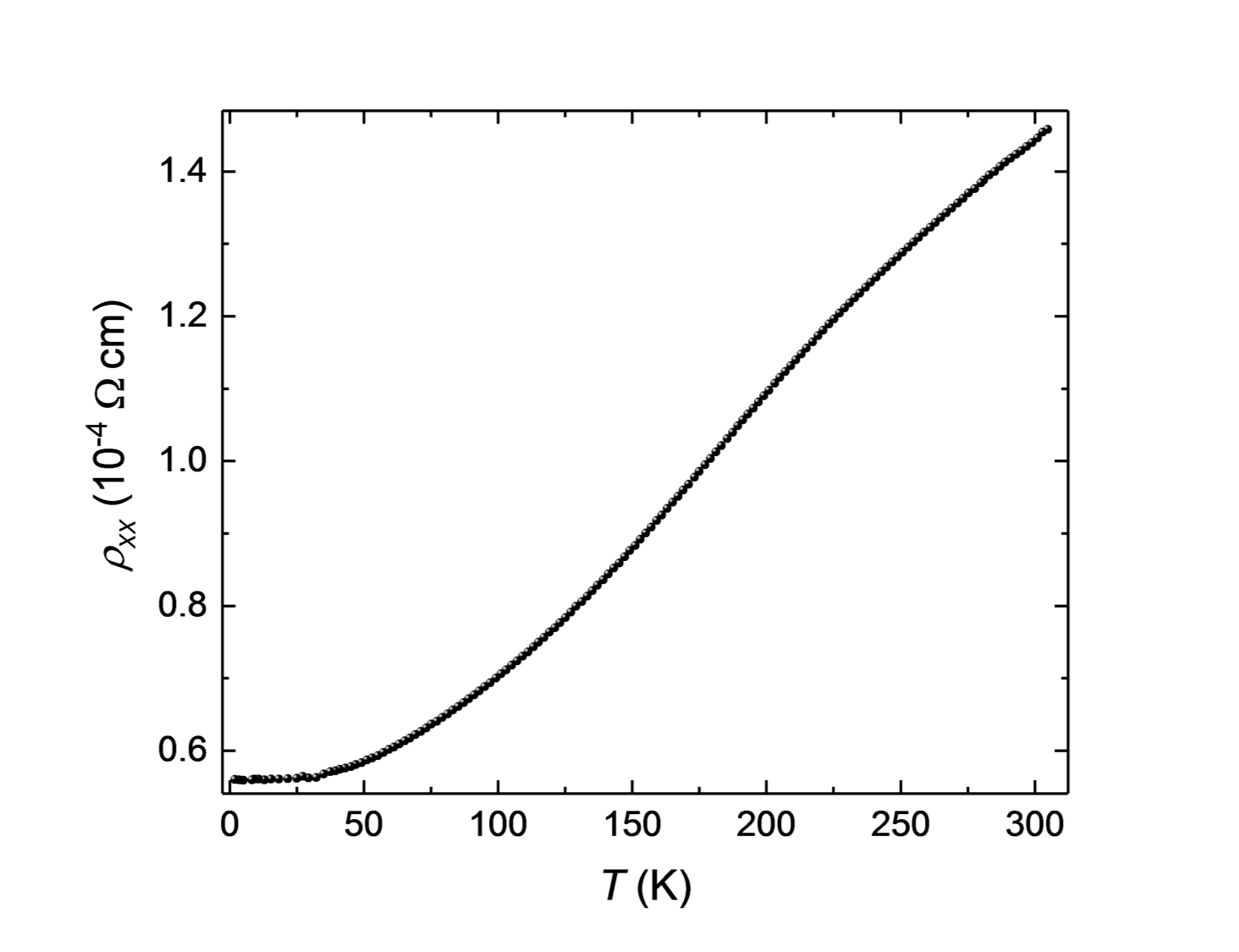}
\caption{\label{FigS3} \textbf{Longitudinal resistivity.} Resistivity as a function of temperature for a Co$_2$MnGa single crystal with current along the [001] direction.}
\end{figure*}

The magnetic hysteresis loop recorded at 2 K shows a soft ferromagnetic behavior, (Fig. \ref{FigS2}\pana). The magnetization saturates above $\sim 0.5$ T field with saturation magnetization $M_\textrm{S} \sim 4.04 \pm 0.11 \mu_{\textrm{B}}$/f.u. Earlier neutron diffraction experiments reported a moment of $3.01 \pm 0.16\ \mu_{\textrm{B}}/$Mn atom and $0.52 \pm 0.08\ \mu_{\textrm{B}}/$Co atom, with negligible moment on Ga and total moment $4.05 \pm 0.05\ \mu_{\textrm{B}}$/f.u. \cite{Co2MnGa_neutron}, consistent with our saturation magnetization measurement. Evidently, the compound follows the Slater-Pauling rule, $M_\textrm{S} = N-24$, where $N$ is the number of valence electrons, $N = 28$ for Co$_2$MnGa. We see a ferromagnetic loop opening with coercive field $\sim 35$ Oe (Fig. \ref{FigS2}\pana, inset). We observe that $M_\textrm{S}$ decreases slightly with increasing temperature (Fig. \ref{FigS2}\panb). An earlier work addressing the origin of the ferromagnetic phase performed a first-principles calculation of exchange interaction parameters in \s\ and found that the leading contribution is provided by exchange between 3$d$ orbitals on nearest-neighbor Mn and Co sites, see Tables II and III in \cite{Co2MnGa_exchange}. This nearest-neighbor Mn-Co exchange was found in calculation to have a positive sign, favoring ferromagnetic alignment. Next, we study the temperature dependent longitudinal resistivity of our samples $\rho_{xx} (T) $, with zero applied magnetic field (Fig. \ref{FigS3}). We measure current along the [100] direction. Clearly the compound shows metallic behavior throughout the temperature range with very low residual resistivity: $\rho_{xx}$ $(2\textrm{K}) \sim 5.6 \times 10^{-5}$ $\Omega$ $\textrm{cm}$ and residual resistivity ratio (RRR) $\rho_{xx}$ $(300 \textrm{K})/\rho_{xx}$ $(2 \textrm{K}) = 2.6$.\\

\begin{figure*}
\centering
\includegraphics[width=16cm,trim={0in 0.2in 0in 0in},clip]{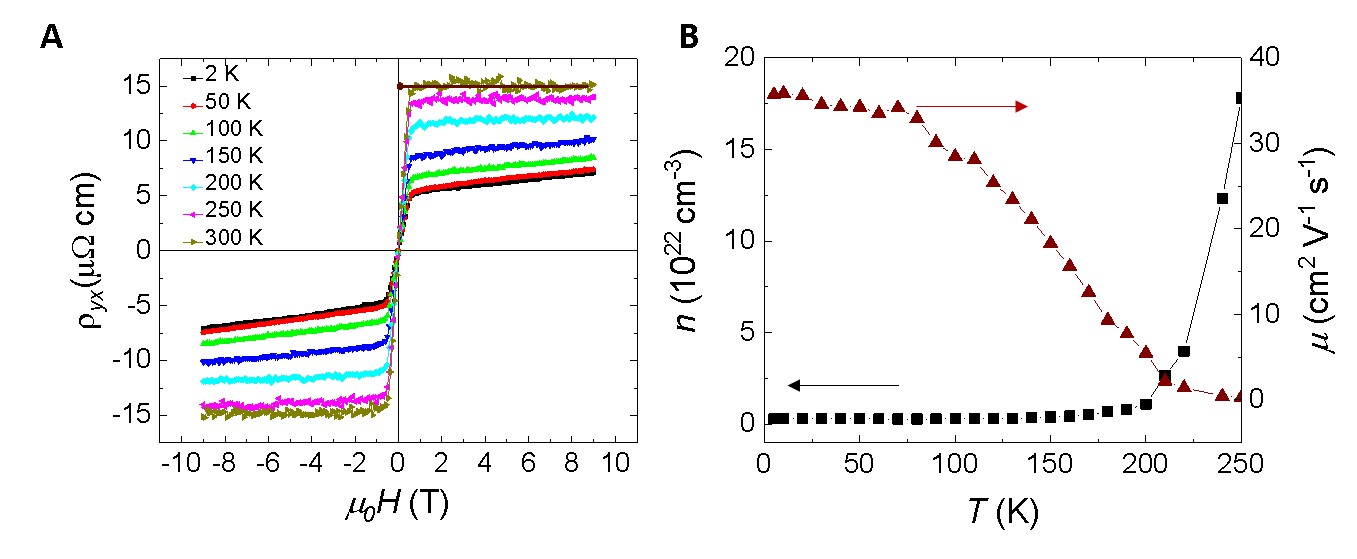}
\caption{\label{Hall} \textbf{Hall resistivity.} (\textbf{\pana}) Magnetic field dependence of the Hall resistivity at several representative temperatures. The magnetic field is applied along the [110] direction and the current along [001]. (\textbf{\panb}) Temperature dependence of carrier concentration and mobility of \s\ calculated from the Hall coefficient of the $\rho_{yx}$ data, as described in the text.}
\end{figure*}

\begin{figure}
\centering
\includegraphics[width=14cm,trim={0in 0in 0in 0in},clip]{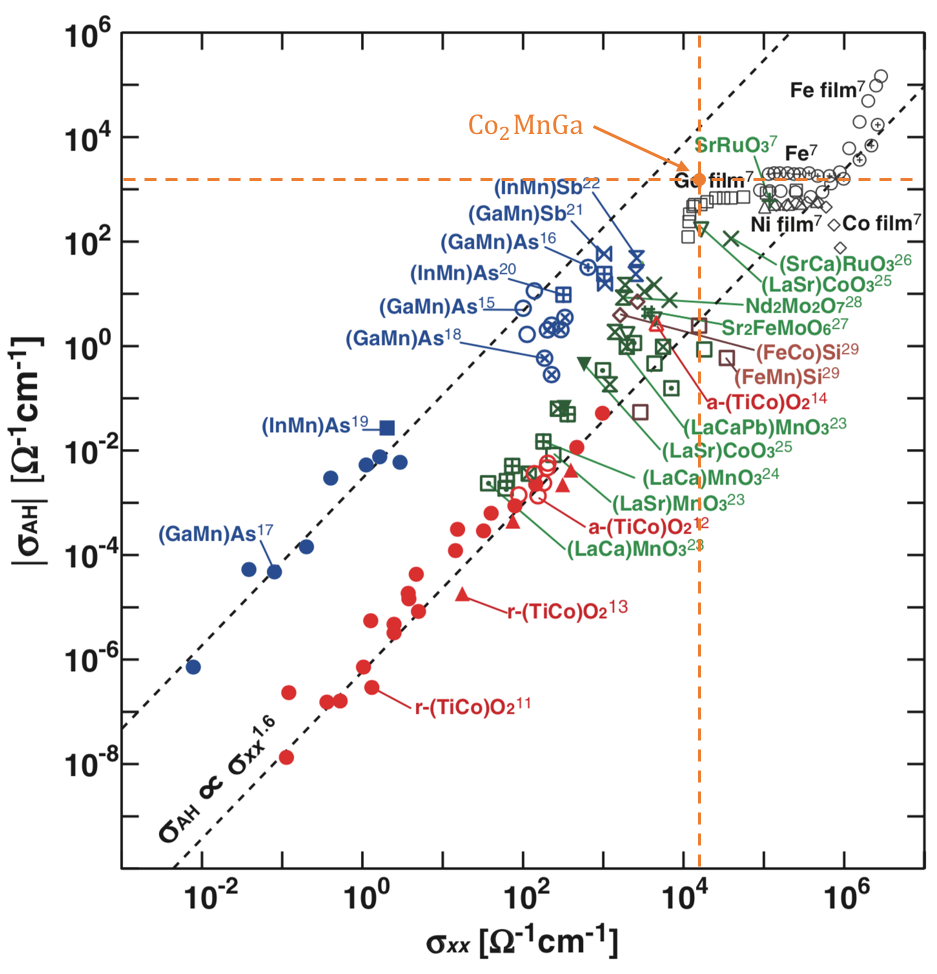}
\caption{\label{AHE_fig} The anomalous Hall response in various materials, reproduced from \href{http://iopscience.iop.org/article/10.1143/JJAP.46.L642}{\textit{Jpn. J. Appl. Phys.} {\bf 46}, L642 (2007)} (see references within). We add the measured values for \s. Only Fe (shown here) and Co$_2$MnAl (Ref. \cite{Co2MnAl_Jakob}) are known to have a larger AHE.}
\end{figure}

We provide some additional background on the Hall measurements presented in maintext Fig. 4. The Hall resistivity $\rho_{xy}$ is generally expressed as,
\beq
\rho_{xy}=R_0 \mu_0 H + \rho_{\textrm{AH}}
\eeq
where $R_0$ is known as the ordinary Hall coefficient arising from the Lorentz force and $\rho_\textrm{AH}$ is the anomalous Hall contribution \cite{ReviewAHE_Nagaosa_Ong}. The Hall conductivity $\sigma_{xy}$ is defined from the matrix inverse of $\rho$,
\beq
\sigma_{xy} = \frac{-\rho_{xy}}{\rho^2_{xy} + \rho^2_{xx}}
\eeq

where $\rho_{xx}$ is the longitudinal resistivity. Therefore, we can obtain $\sigma_{\textrm{AH}}$ (or, in a similar way, $\rho_{\textrm{AH}}$) by extrapolating the high field $\sigma_{xy}$ value back to $\mu_0 H = 0$ to find the $y$ intercept, as shown in main text Fig. 4\pana\ and Fig. \ref{Hall}\pana. We find $\sigma_{\textrm{AH}} = 1530 \ \Omega^{-1} \ \textrm{cm}^{-1}$ at 2 K, the largest anomalous Hall response in any known material except Fe and Co$_2$MnAl, Fig. \ref{AHE_fig} and Ref. \cite{Co2MnAl_Jakob}. Further, from the high field slope of $\rho_{yx}$ as a function of $\mu_0 H$, we find an ordinary Hall coefficient $R_0 = 2.76 \times 10^{-3} \ \textrm{cm}^3/\textrm{C}$ at 2 K and $9.98 \times 10^{-5} \ \textrm{cm}^3/\textrm{C}$ at 300 K. The positive sign of $R_0$ suggests that the charge carriers in \s\ are majority hole type through the full temperature range. We estimate the carrier concentration as $n=1/(eR_0)$ and the carrier mobility as $\mu = R_0/\rho_{xx}$, where $e$ is the electron charge, Fig. \ref{Hall}\panb.

\subsection{Survey band structure calculation}

We consider a bird's eye view of the \ai\ bulk band structure in the ferromagnetic state (Fig. \ref{FigSurvey}). We observe two majority spin bands near the Fermi level---these are the bands which form the Weyl lines \cite{Co2MnGa_Guoqing}. There is also a large, irrelevant minority spin pocket around $\Gamma$ which we experimentally suppress by judicious choice of photon energy in ARPES. For completeness, we present ARPES measurements on this minority spin pocket below (Fig. \ref{spinmin}). Without magnet order, the \ai\ band structure changes drastically (Fig. \ref{FigNonMag}). This provides additional evidence that in ARPES we access the magnetic state of \s. Moreover, this result suggests that the Weyl lines we observe are essentially magnetic in the sense that they disappear if we remove the magnetic order.\\

We also perform a band structure calculation taking into account spin-orbit coupling (SOC) and we find that the gap opened is of order $\sim$ meV, negligible for our ARPES measurements (Fig. \ref{SOC_calc}).

\subsection{Weyl lines from calculation}

Here we give a more systematic introduction to the full network of Weyl lines in \s\ as predicted by \ai\ \cite{Co2MnGa_Guoqing}. Recall that a Weyl line is a one-dimensional crossing between a pair of singly-degenerate bands. In \s, the Weyl lines are contained in the mirror planes of the bulk Brillouin zone but they are allowed to disperse in energy. As a result, it is instructive to plot each Weyl line as a function of $k_x, k_y$ and $E_\textrm{B}$, where $k_x$ and $k_y$ without loss of generality are the two momentum axes of the mirror plane. \textit{Ab initio} predicts three independent Weyl lines in Co$_2$MnGa, which we denote the red, blue and yellow Weyl lines (Fig. \ref{FigCalc}\panb-\pand). The energies are marked with respect to the Fermi level observed in numerics. Since we find in experiment that the Fermi level is at $-0.08 \pm 0.01$ eV relative to calculation, the experimental Fermi level cuts through all of the Weyl lines. To better view the full pattern of line nodes throughout the bulk Brillouin zone, we collapse the energy axis and plot the line nodes in $k_x, k_y, k_z$ (Fig. \ref{FigCalc}\pana). Although we start with three independent line nodes, these are each copied many times by the symmetries of the crystal lattice, giving rise to a rich line node network.\\

Since we are studying the (001) surface, it's useful to consider more carefully how the line nodes project onto the surface Brillouin zone---in other words, to see how Fig. 2\panf\ arises from Fig. \ref{FigCalc}\pana. To start, the center red line node will project straight up on its face around $\bar{\Gamma}$. By contrast, the adjacent blue line nodes are ``standing up'' and will project on their side, so that in fact there will be a ``double'' cone in the surface projection. Moreover, the blue line node projection will form an open line node segment. The other red line nodes will also project on their side, forming ``double" red cones along an open line segment. The yellow line nodes also projects in two distinct ways. The yellow line nodes in the $k_z = 0$ plane will produce a single yellow line node projection, while the yellow line nodes in the $k_x = 0$ and $k_y = 0$ planes are standing up, so they will produce a double yellow line node in an open segment. Band folding associated with the (001) surface projection further sends the $k_z = 0$ yellow line node towards $\bar{\Gamma}$.

\subsection{ARPES systematics on the blue Weyl line}

We present an extended dataset for main text Fig. 1\pand-\panh\ (Fig. \ref{BlueFS}), with a finer \eb\ sampling. We observe in greater detail the evolution of the dispersion from a $<$ to a dot to a $>$. For instance, we find that the crossing point moves systematically away from $\bar{\Gamma}$ with deeper \eb, consistent with a line node. In \ai, we observe a similar evolution for the blue Weyl line (Fig. \ref{BlueFSCalc}\pana,\panb). We can better understand this evolution by considering the constant-energy surfaces for a generic line node (Fig. \ref{BlueFSCalc}\panc,\pand). For \eb\ above the line node, the slice intersects only the upper cone, giving {\bf I}. For \eb\ which cross the line node, we find electron and hole pockets intersecting at a point, as in {\bf II}. As we continue to move downward the intersection point traces out the line node, shifting from left to right. Comparing to our ARPES spectra, we observe that the photoemission cross-section appears to be dominated by the intersection point for this range of \eb. Lastly, as we scan below the line node, the intersection point completely zips closed the electron pocket and zips open the hole pocket, as in {\bf III}. A detailed study of the $E_\textrm{B}$ dependence of the constant energy surface in ARPES again suggests a line node in \s.\\

To provide another perspective, we cut parallel to the blue Weyl line (Fig. \ref{FigBluePar}). In contrast to main text Figs. 1,2, here we cut \textit{along} the Weyl line. Sweeping in $k_y$, we see the conduction and valence bands approach (Fig. \ref{FigBluePar}\pana-\pand), touch each other at fixed $k_y^\textrm{L} = - 0.03\ \textrm{\AA}^{-1}$ along a finite range of $k_x$ (Fig. \ref{FigBluePar}\pane) and then move apart again (Fig. \ref{FigBluePar}\panf-\pani). These parallel $E_\textrm{B} - k_x$ cuts again suggest a line node dispersion.\\

Next, we perform a Lorentzian peak fitting of the blue Weyl line. We begin with the $E_\textrm{B}-k_x$ cuts discussed in main text Fig. 2 and we choose the energy distribution curve (EDC) passing through the crossing point (Fig. \ref{FigBlueFit}). We fit these EDCs to the following form,

\beq
I(x) = (C + L_1(x) + L_2(x))f(x)
\eeq

\beq
L_i(x) = \frac{A_i^2}{(x - B_i)^2 + C_i^2}  \hspace{1cm}    f(x) = (\exp(\beta(x - \mu)) + 1)^{-1}
\eeq\\
We include two Lorentzian peaks $L_1(x)$ and $L_2(x)$, where the first peak corresponds to the line node crossing LN, while the second peak corresponds to a deeper valence band VB' which is useful for improving the fit. We also include the Fermi-Dirac distribution $f(x)$ and a constant offset $C$ which we interpret as a background spectral weight approximately constant within the energy range of the fit. We find a high-quality fit close to $\bar{\Gamma}$ (Fig. \ref{FigBlueFit}\pana-\pand) using a single LN peak. Away from the crossing point, the peaks are well-described by a linear dispersion, further suggesting a band crossing (Fig. \ref{FigBlueFit}\panc). At $k_y = 0.45$ \invA, we observe that the fit begins to deviate from the data, and at $k_y = 0.5$ \invA there is an even more noticeable error (Fig. \ref{FigBlueFit}\pane-\panh). We speculate that this deviation may arise due to our finite $k_y$ resolution/linewidth as well as the fact that these spectra cut near the extremum of the Weyl line, producing a smeared-out energy gap. Another explanation considers the detailed dispersion of the blue Weyl line, which exhibits a rapid upward dispersion at its extremum (Fig. \ref{FigCalc}\panc). Due to broadening along $k_y$, we may capture LN peaks from a range of $k_y$, smearing out this rapid dispersion and producing a plateau structure in the EDC. For $k_y = 0.54$ \invA, we clearly observe two peaks on the EDC, so we fit with an additional Lorentzian $L_3 (x)$ (Fig. \ref{FigBlueFit}\pani-\panj). This gives VB and CB, the peaks corresponding to the conduction and valence bands of the line node. This interpretation is consistent with \ai, which predicts that the blue Weyl line ends at $k_y = 0.5$ \invA.\\

Lastly, we take the results of our peak fitting and compare them with the calculated blue line node dispersion. We plot the LN peak maxima and the standard deviation of the peak positions, Fig. \ref{FigBlueTrack}. We ignore EDCs at $k_y > 0.45$ \invA\ because the plateau shape in the EDC is poorly described by a single Lorentzian, as discussed above. We compare these numerical fitting results with a first-principles calculation of the blue Weyl line dispersion, shifted by $0.08$ eV. We find a reasonable quantitative agreement between the fit and calculated dispersion. Note that there is some expected contribution to the error from the \textit{ab initio} calculation as well as corrections to the Lorentzian fitting form. These results support our observation of a line node in our ARPES spectra on Co$_2$MnGa.

\subsection{ARPES study of the yellow Weyl line}

We can also observe signatures of yellow Weyl lines in our ARPES data, at incident photon energy $h\nu = 50$. We can identify a candidate yellow Weyl line by comparing an ARPES constant-energy surface and the projected nodal lines (Fig. \ref{FigYellow1}\pana, \panb). We reiterate that there are two different ways in which the yellow Weyl lines can project on the (001) surface. In particular, the four yellow Weyl lines along $\bar{\Gamma} - \bar{M}$ are ``standing up'', so two crossings project onto the same point in the surface Brillouin zone, similar to the blue Weyl line we discussed above. By contrast, the outer yellow Weyl line runs in a single large loop around the entire surface Brillouin zone. It projects ``lying down'', with single crossing projections. Here we focus on the double yellow line node. We study constant-energy surfaces at various binding energies (Fig. \ref{FigYellow1}\panc-\pane) and we observe the same $<$ to $>$ switch that we discussed in the case of the blue Weyl line. We see similar behavior on an \textit{ab initio} constant energy surface (Fig. \ref{FigYellow1}\pani-\pank). Note crucially that the electron-to-hole transition occurs in the same direction in the ARPES spectra and calculation, suggesting that the line node dispersion has the same slope in experiment and theory. This provides additional evidence for the yellow Weyl line. Next we search for signatures of the yellow Weyl line on a series of $E_\textrm{B}-k_a$ cuts (green lines in Fig. \ref{FigYellow2}\pang, \panh). We observe the upper cone near the center of the cut for $k_a$ closest to $\bar{\Gamma}$ (Fig. \ref{FigYellow2}\pana), corresponding to the yellow markers in Fig. \ref{FigYellow1}\panc. As we slide away from $\bar{\Gamma}$, we observe the band crossing and lower cone (Fig. \ref{FigYellow2}\panb, \panc). To pinpoint the cone, we study MDCs through the line node. We observe twin peaks corresponding to the upper and lower cone (\ref{FigYellow2}\pand,\panf), as well as a single peak when we cut through the line node (\ref{FigYellow2}\pane). In this way, we observe signatures of the ``double'' yellow Weyl line of Co$_2$MnGa in our ARPES spectra.\\

Lastly, we search for signatures of the other, ``single'' yellow Weyl line. The outer features in Fig. \ref{FigYellow2}\panb, \panc, as well as the large off-center peaks in Fig. \ref{FigYellow2}\pane, \panf, correspond well to the predicted locations of the single yellow Weyl line. The valence band further shows a cone shape. However, we note that the conduction band appears to be shifted/offset in momentum relative to the valence band, for instance near the Fermi level at $k_a \sim 0.25$ $\textrm{\AA}^{-1}$ (Fig. \ref{FigYellow2}\panb). This suggests that we are perhaps observing a surface state or surface resonance which partly traces out the line node. This explanation appears to be consistent with our calculations, compare main text Fig. 3\pane, which show a similar surface resonance. In summary, we provide evidence for the yellow Weyl lines in our ARPES spectra.

\subsection{ARPES study of the red Weyl line}

Having discussed the blue and yellow Weyl lines, we search for ARPES signatures of the red Weyl line. On the constant-energy surfaces, we observe a square feature around $\bar{\Gamma}$ (Fig. \ref{FigRed}\panf-\panh). This corresponds well to the predicted red Weyl line (Fig. \ref{FigRed}\pand). Next, we study a series of $E_\textrm{B}-k_x$ cuts passing through the center square feature. We see two clear branches dispersing away from $k_x = 0\ \textrm{\AA}^{-1}$ as we approach the Fermi level (red arrows, Fig. \ref{FigRed}\pana-\panc). We can mark these features on an MDC (Fig. \ref{FigRed}\pane). However, we observe no cone or crossing. We speculate that this may be because the other branch of the Weyl line has low photoemission cross-section under these measurement conditions.

\subsection{Photon energy dependence of the drumhead surface state}

We present an extended dataset to accompany main text Fig. 3, showing the drumhead surface states. In main text Fig. 3, we presented an energy distribution curve (EDC) stack cutting through the drumhead surface state at different photon energies. Here we present the full $E_\textrm{B}-k_x$ cut for each photon energy included in the stack. We observe the drumhead surface state consistently at all energies (orange arrow, Fig. \ref{DH1}\pana-\pani). Additionally, we show an EDC stack at a different momentum, $k_{||} = 0.45 \ \textrm{\AA}^{-1}$, which cuts not through the drumhead surface state but the yellow line node cone (Fig. \ref{DH1}\panj) analogous to main text Fig. 3\pang. When cutting through the candidate drumhead, the EDC stack showed no dispersion in the peak energy as a function of photon energy, indicating no $k_z$ dispersion and suggesting a surface state. Here, by contrast, we can observe that the peak positions shift with photon energy (blue arrows). This shift suggests that the yellow line node lives in the bulk.

\subsection{In-plane dispersion of the drumhead surface state}

We briefly study the in-plane dispersion of the drumhead surface states. At $h\nu = 35$ eV, with Fermi surface as shown in Fig. \ref{DH2}\pana, we study a sequence of $E_\textrm{B}-k_x$ cuts scanning through the drumhead surface state, Fig. \ref{DH2}\panb-\pane. We observe that the surface state disperses slightly downward in energy as we scan away from $\bar{\Gamma}$ and that it narrows in $k_{||}$, as expected because the line node cones move towards each other as we approach the corner of the Brillouin zone. The dispersion of the candidate drumhead in-plane (but not out-of-plane) is consistent with the behavior of a surface state.

\subsection{ARPES and \ai\ study of the minority spin pocket}

\begin{figure*}
\centering
\includegraphics[width=12cm,trim={0in 0.5in 0in 0in},clip]{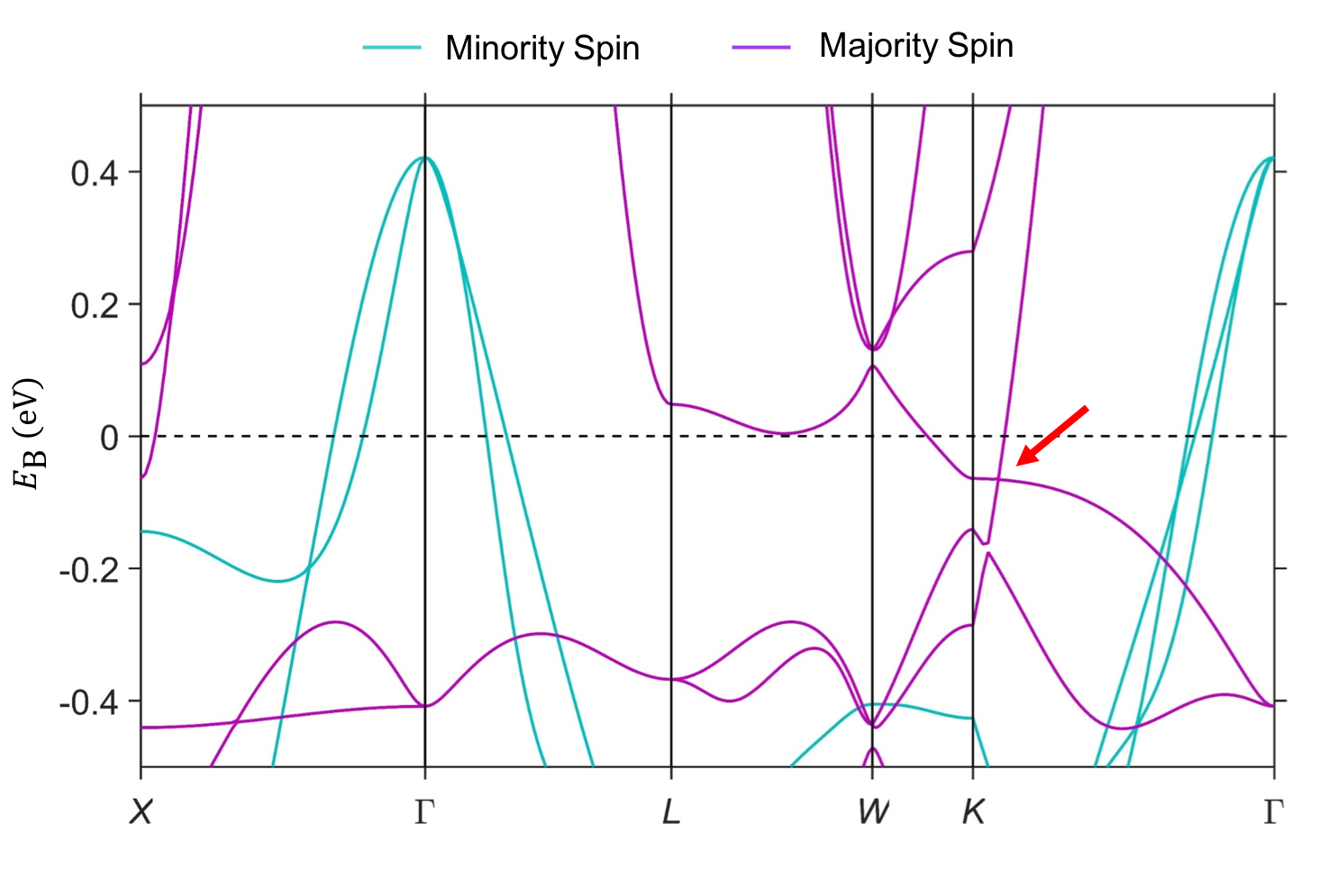}
\caption{\label{FigSurvey} \textbf{Survey band structure of Co$_2$MnGa.} \textit{Ab initio} band structure of \s\ in the ferromagnetic state. Two majority spin bands near the Fermi level form Weyl lines (orange arrow).}
\end{figure*}

\begin{figure*}
\centering
\includegraphics[width=13cm,trim={0in 0in 0in 0in},clip]{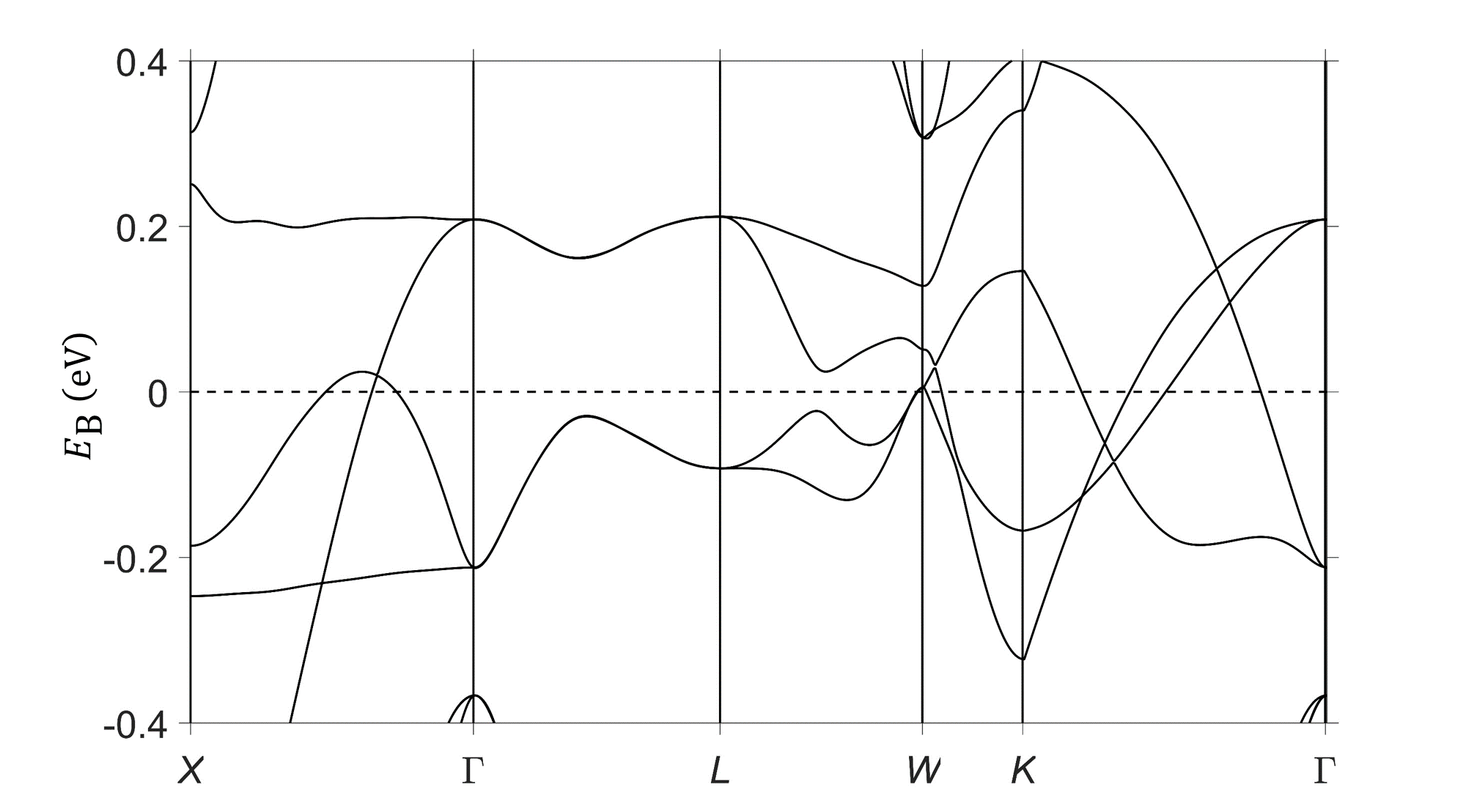}
\caption{\label{FigNonMag} \textbf{Non-magnetic survey band structure.} \textit{Ab initio} band structure, ignoring ferromagnetism. The magnetic Weyl lines disappear.}
\end{figure*}

\begin{figure*}
\centering
\includegraphics[width=14cm,trim={0in 0.02in 0.02in 0in},clip]{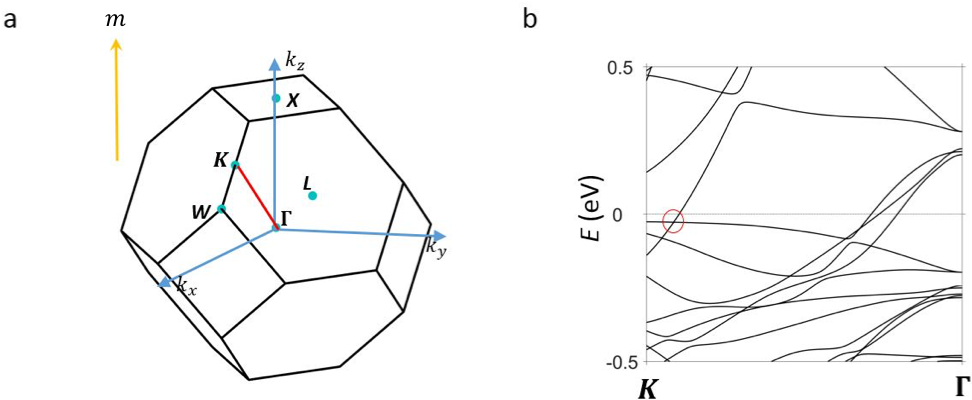}
\caption{\label{SOC_calc} \textbf{Band structure with spin-orbit coupling (SOC) in \s.} We observe numerically that, cutting along the $\Gamma-K$ direction in the bulk Brillouin zone, the band gap opened in the magnetic Weyl line is $< 1$ meV. The yellow arrow represents the magnetization direction $m$.}
\end{figure*}

\begin{figure*}
\centering
\includegraphics[width=16cm,trim={1in 5.4in 1in 1.6in},clip]{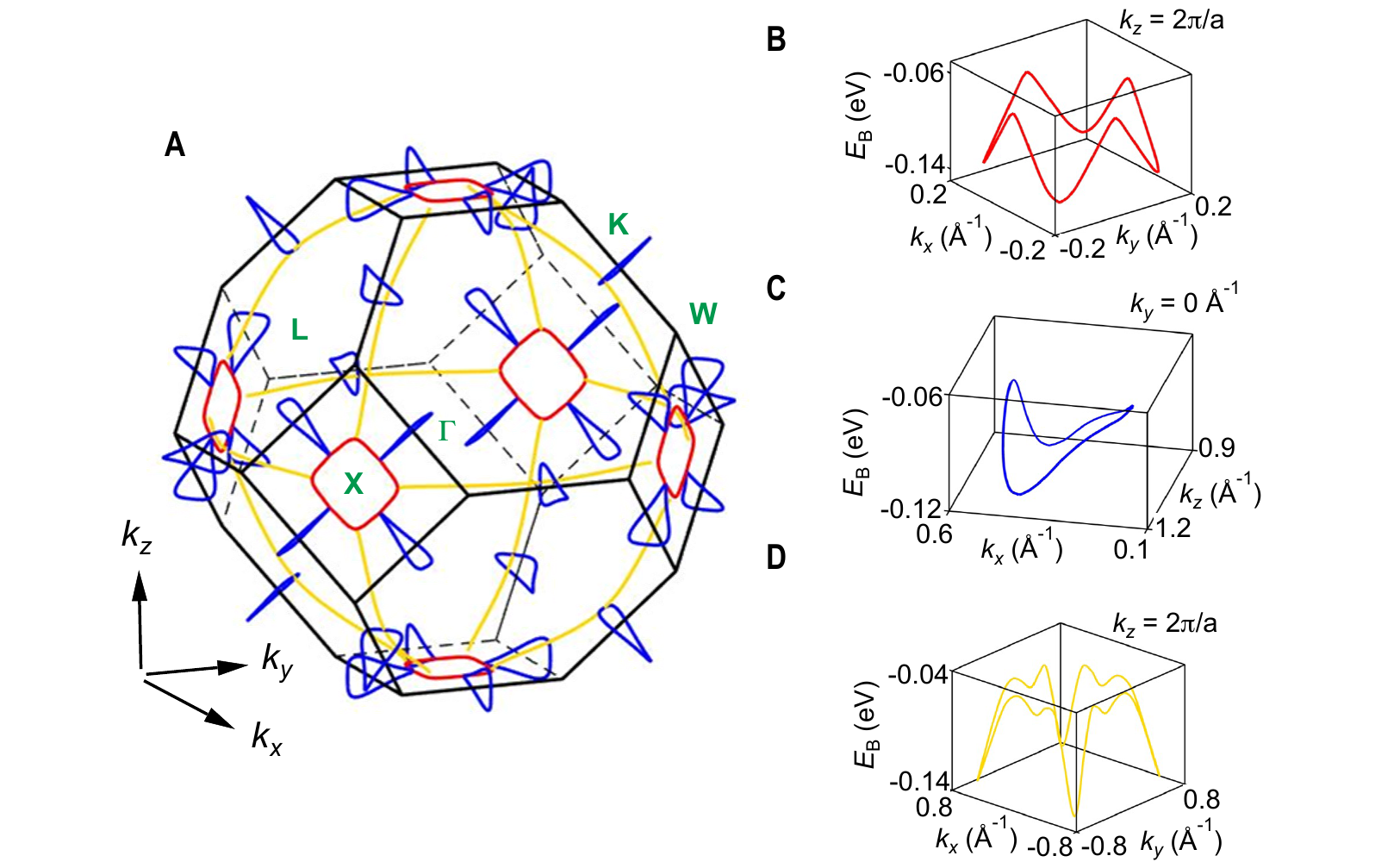}
\caption{\label{FigCalc} \textbf{Predicted Weyl lines in Co$_2$MnGa.} (\textbf{\pana}) \textit{Ab initio} prediction of Weyl lines in Co$_2$MnGa, with the $E_\textrm{B}$ axis collapsed. (\textbf{\panb}-\textbf{\pand}) Theory predicts three unique Weyl lines, which we call red, blue and yellow. Although they are locked on a mirror plane, the line nodes disperse in energy. Each one is copied many times through the Brillouin zone by the symmetries of the lattice.}
\end{figure*}

\begin{figure*}
\centering
\includegraphics[width=16cm,trim={1.4in 4.8in 1.4in 1.1in},clip]{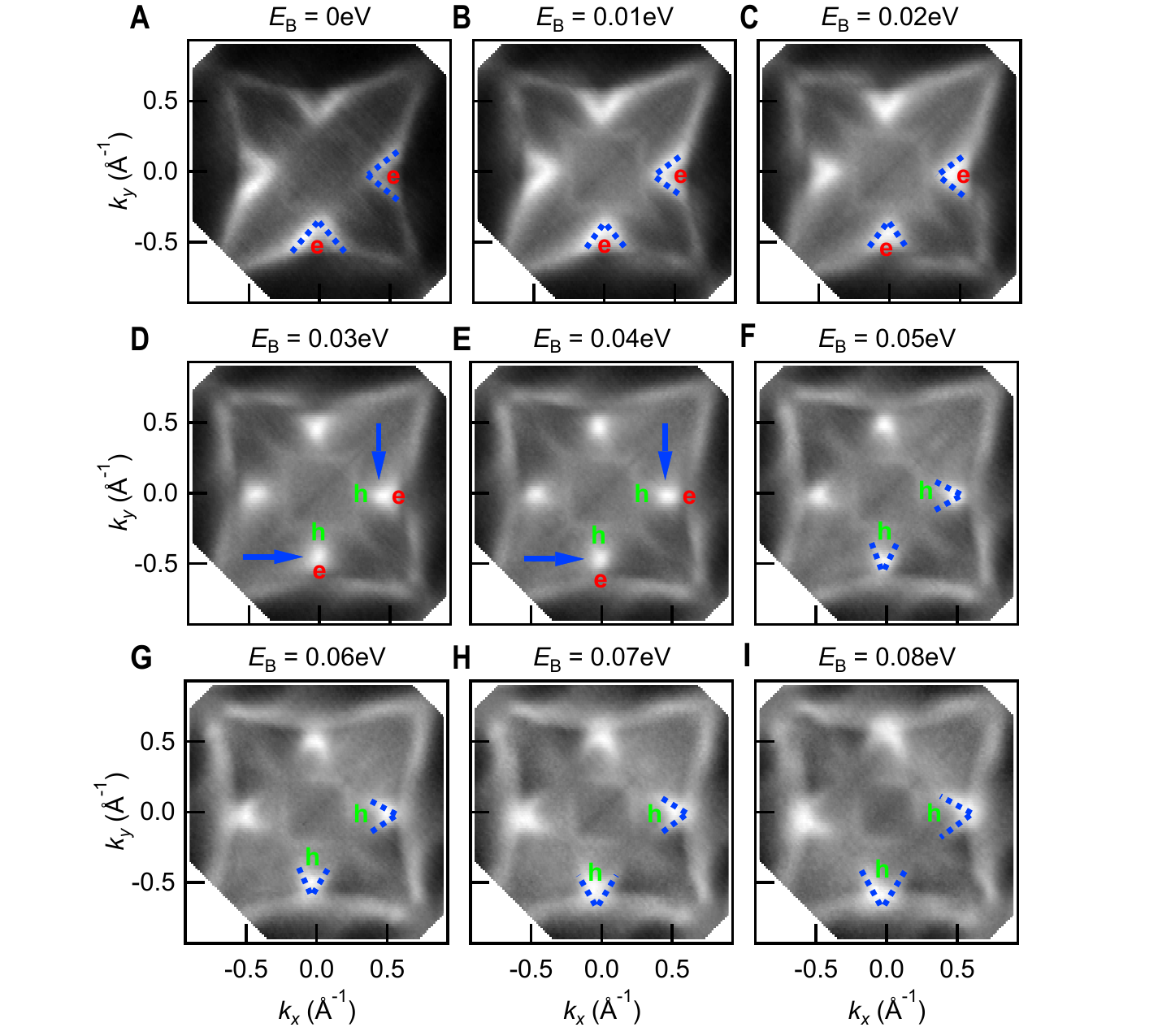}
\caption{\label{BlueFS} \textbf{Constant energy surfaces of Co$_2$MnGa.} (\textbf{\pana}-\textbf{\pani}) We expand on the dataset shown in main text Fig. 1\pand-\panh\ by plotting constant energy surfaces at additional \eb. The evolution is characteristic of a line node dispersion.}
\end{figure*}

\begin{figure*}
\centering
\includegraphics[width=16cm,trim={1in 6.6in 1in 1.2in},clip]{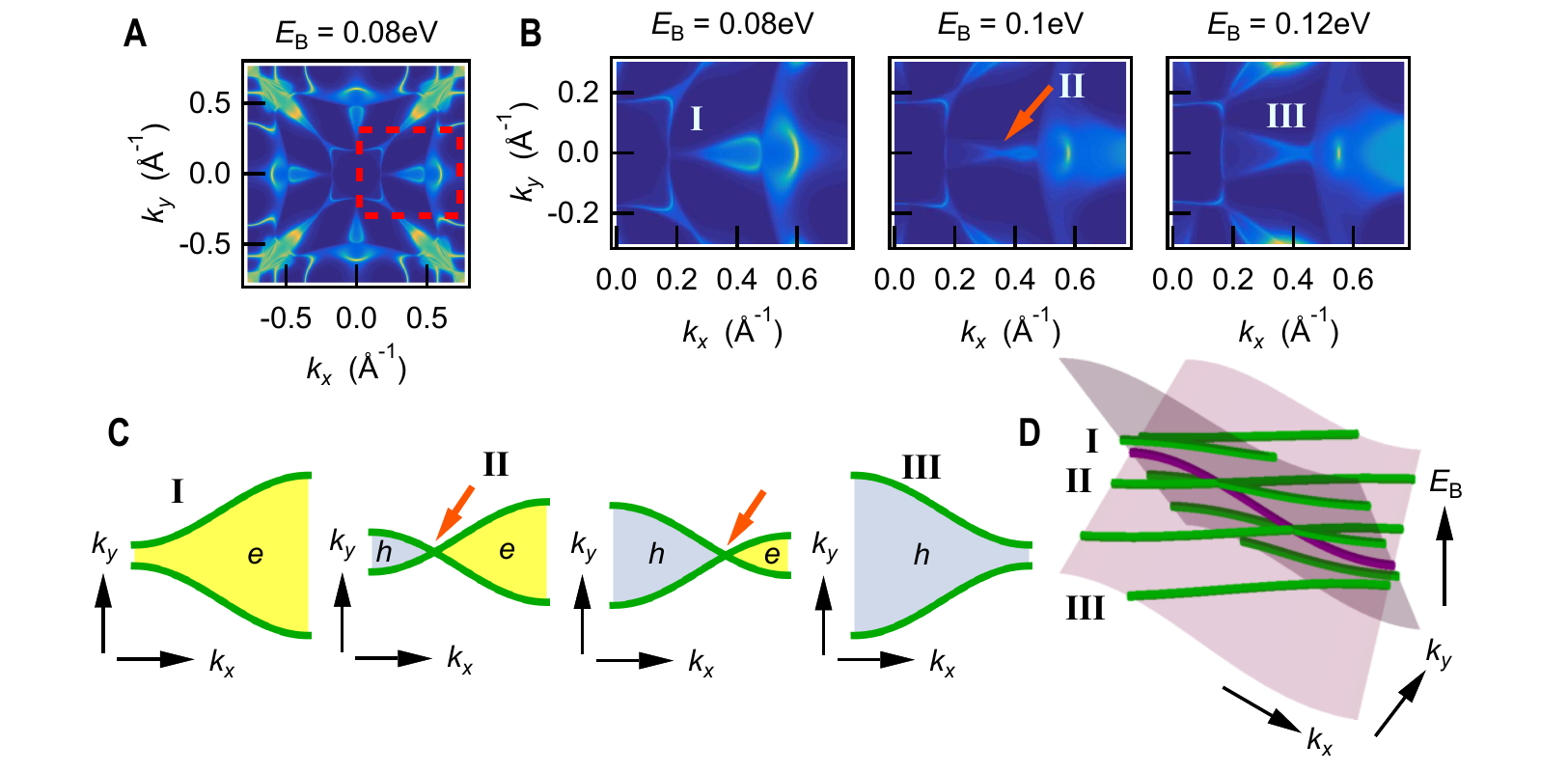}
\caption{\label{BlueFSCalc} {\bf \Ai\ evolution of the blue Weyl line in \eb.} ({\bf \pana}) Constant-energy surface with red boxed region marking ({\bf \panb}) the zoomed-in slices at different \eb. ({\bf \panc}, {\bf \pand}) Cartoon of the constant-energy cuts through a generic line node. Scanning from shallower to deeper \eb, the bands evolve from {\bf I}, an electron pocket, to {\bf II}, a hole and electron pocket touching at a point (orange arrows), to {\bf III}, a hole pocket. The touching point at each energy is a point on the Weyl line.}
\end{figure*}

\begin{figure*}
\centering
\includegraphics[width=16cm,trim={1.2in 4.8in 1.2in 1in},clip]{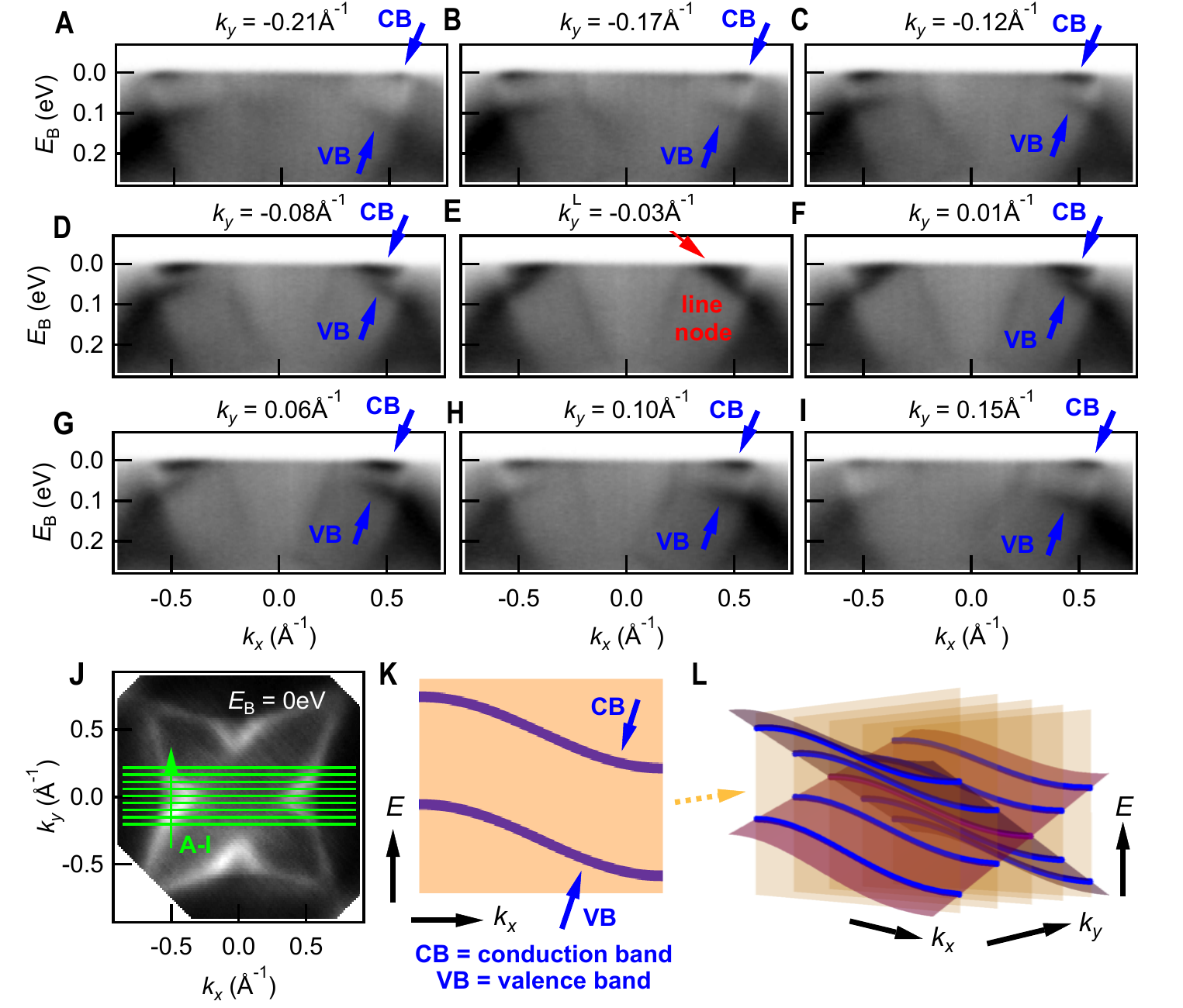}
\caption{\label{FigBluePar} \textbf{Cutting parallel to the bulk Weyl line.} (\textbf{\pana}-\textbf{\pani}) $E_\textrm{B}-k_x$ cuts sweeping through the Weyl line in $k_y$, as indicated in (\textbf{\panj}). The valence and conduction bands appear to meet at a single $k_y^\textrm{L}$ (red arrow). Note that $k_y^\textrm{L}$ is slightly away from $k_y = 0$ $\textrm{\AA}^{-1}$, probably due to a small misalignment. (\textbf{\pank}) A generic cut parallel to a Weyl line and (\textbf{\panl}) its evolution in $k_y$.}
\end{figure*}

\begin{figure*}
\centering
\includegraphics[width=15cm,trim={1.2in 2.0in 1.2in 1.0in},clip]{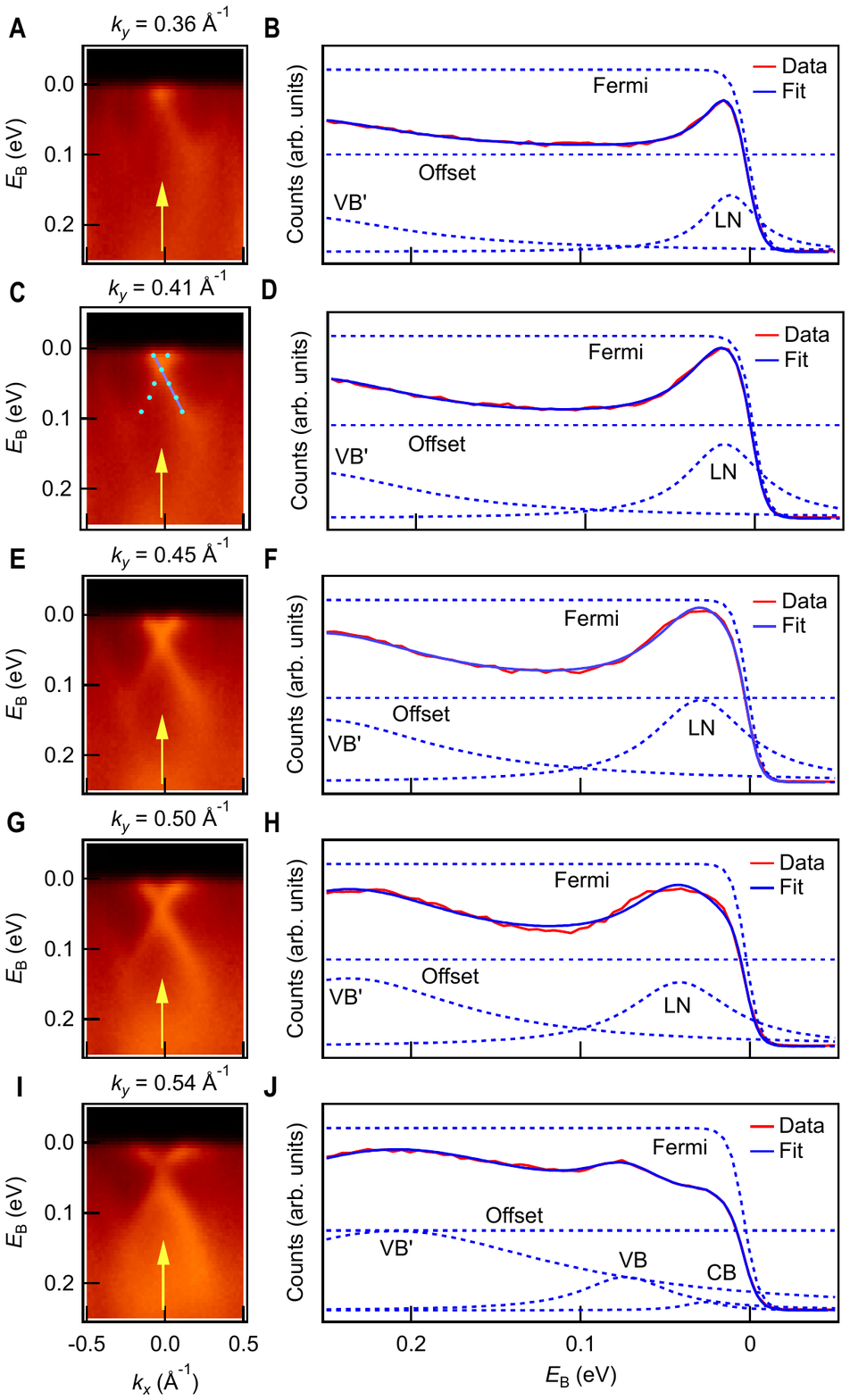}
\caption{\label{FigBlueFit} \textbf{Lorentzian fitting of the blue Weyl line.} (\textbf{\pana}, \textbf{\panc}, \textbf{\pane}, \textbf{\pang}) $E_\textrm{B}-k_x$ cuts through the line node, same  as those in main text Fig. 2\pana. The yellow arrows mark the center EDC. Band dispersion obtained from analysis of the spectra (cyan dots, \panc) with linear fit (purple line, \panc). (\textbf{\pani}) Additional cut further away from $\bar{\Gamma}$, past the predicted end of the blue Weyl line. (\textbf{\panb}, \textbf{\pand}, \textbf{\panf}, \textbf{\panh}, \textbf{\panj}) Fits of the center EDC (fitting form discussed in the text), suggesting a band crossing.}
\end{figure*}

\begin{figure*}
\centering
\includegraphics[width=15cm,trim={1.2in 7in 1.2in 1in},clip]{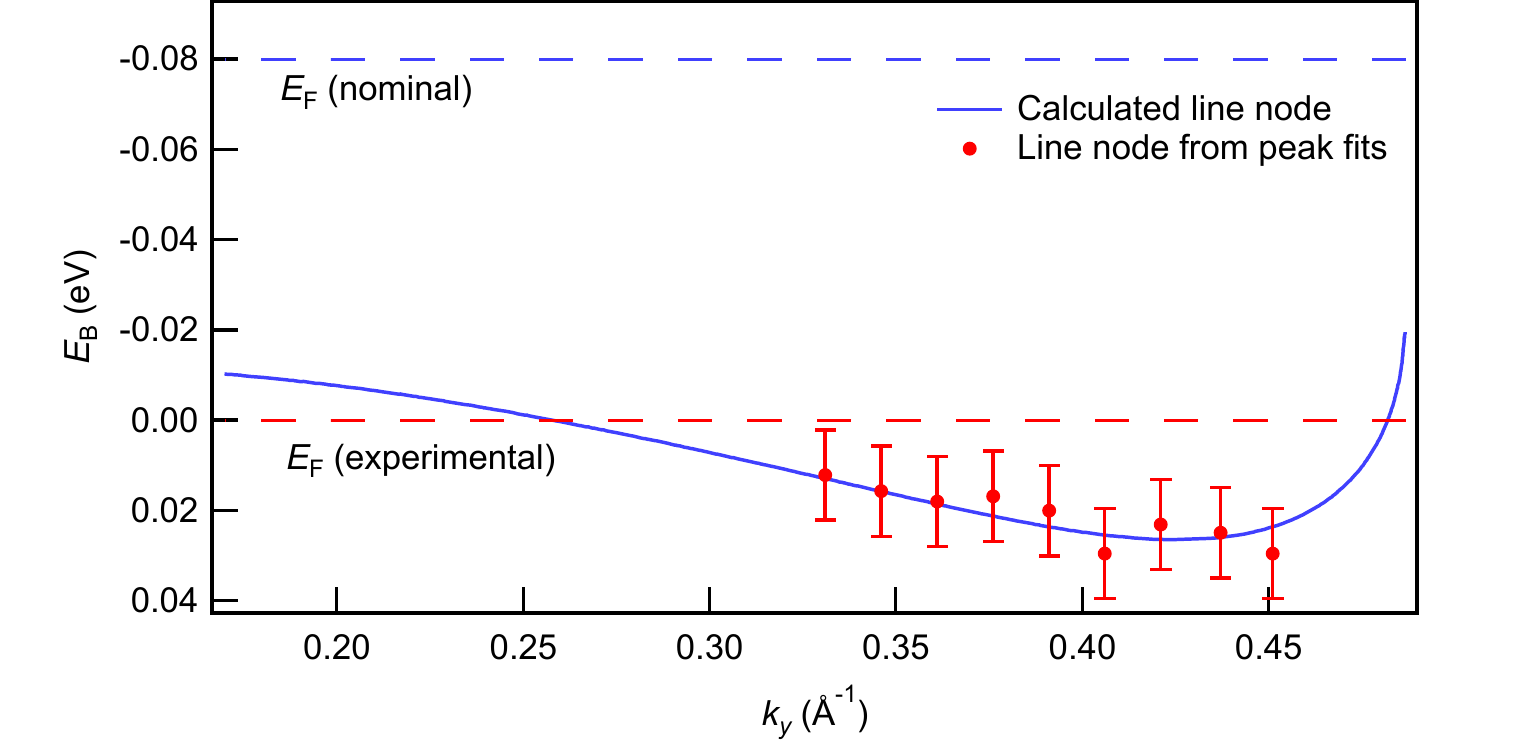}
\caption{\label{FigBlueTrack} \textbf{Comparing Lorentzian peaks to the calculated Weyl line.} Line node dispersion as obtained from Lorentzian peak fits, superimposed on the calculated Weyl line with a $0.08$ eV experimental shift. We observe a match between Lorentzian fits and \textit{ab initio}. The error bars reflect the experimental energy resolution $\delta E = 0.02$ eV.}
\end{figure*}

\begin{figure*}
\centering
\includegraphics[width=16cm,trim={1.2in 4.2in 1.2in 1in},clip]{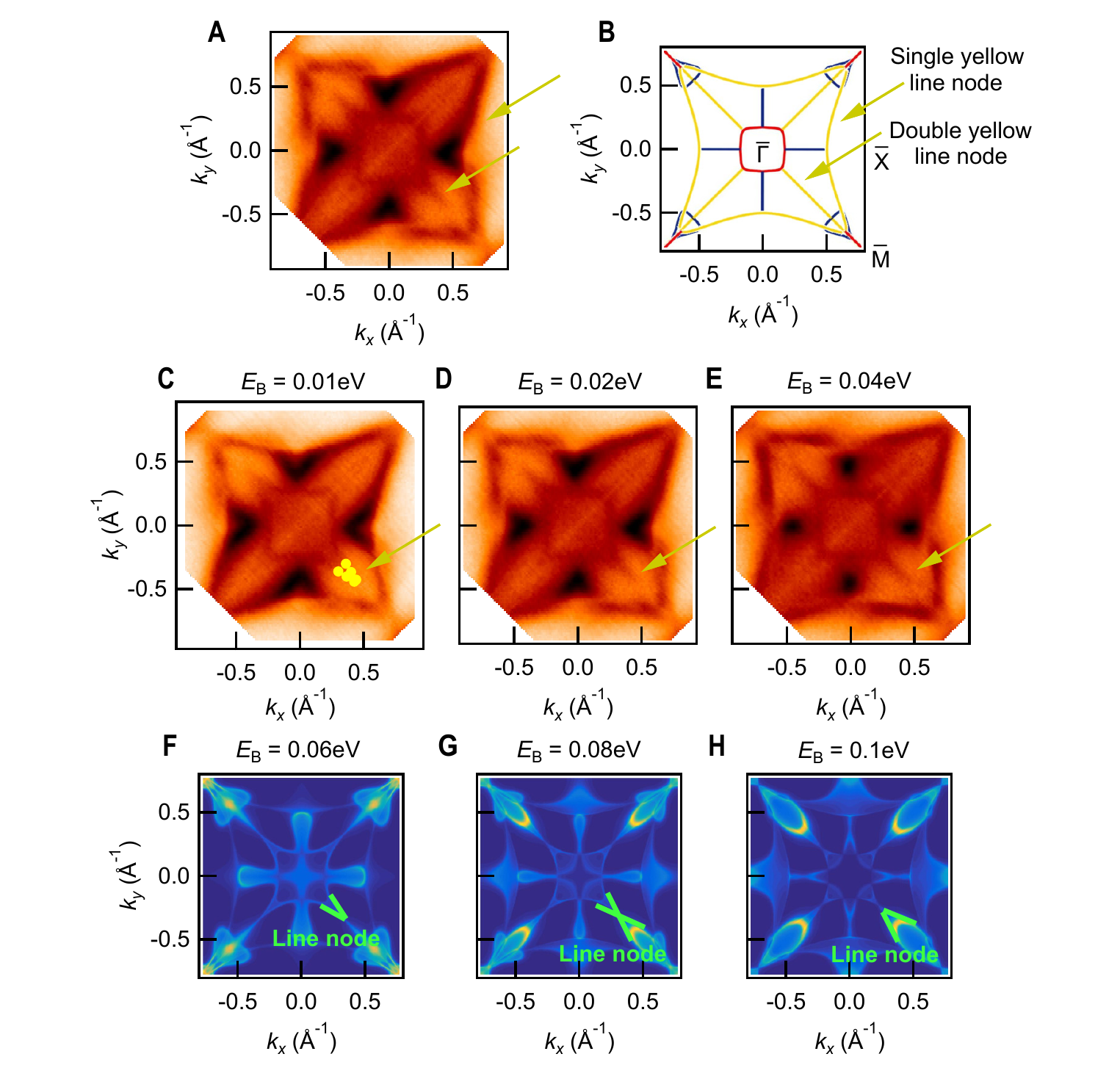}
\caption{\label{FigYellow1} \textbf{Signatures of the yellow Weyl line.} ({\bf \pana}) Constant-energy surface from ARPES compared with ({\bf \panb}) the (001) projection of the line nodes from calculation, emphasizing signatures of the yellow Weyl lines (yellow arrows). Note that the Weyl lines project both ``standing up'' so that the band crossings project in pairs into the surface Brillouin zone (double yellow Weyl line). The remaining yellow line node forms a large ring around the entire surface Brillouin zone, projecting simply ``face up'' (single yellow Weyl line). (\textbf{\panc}-\textbf{\pane}) Constant-energy surfaces showing the characteristic $<$ to $>$ transition (yellow arrows). (\textbf{\panf}-\textbf{\panh}) Corresponding \ai\ cuts showing the double yellow Weyl line (green guides to the eye).}
\end{figure*}

\begin{figure*}
\centering
\includegraphics[width=16cm,trim={1.2in 3.1in 1.2in 1.3in},clip]{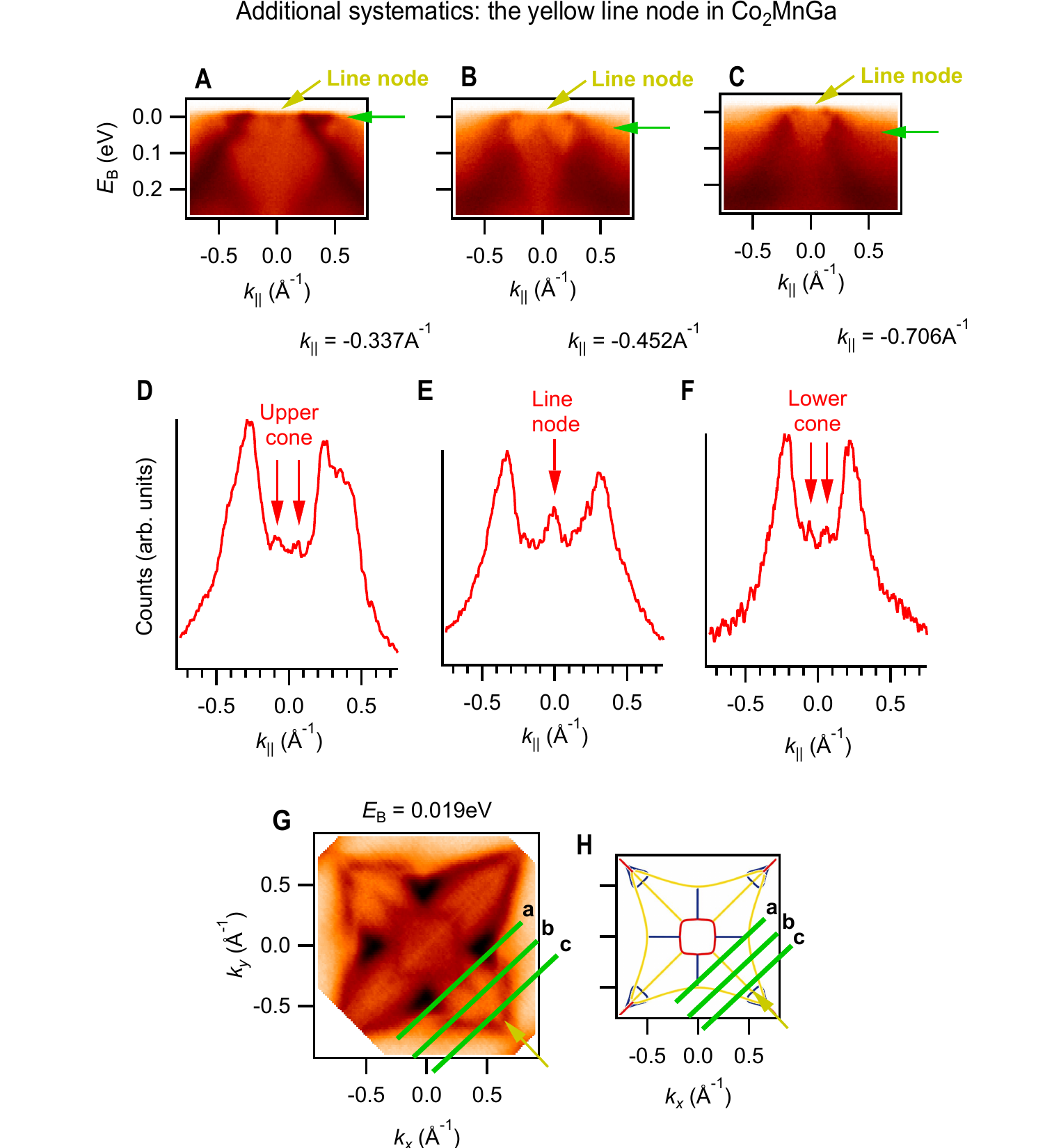}
\caption{\label{FigYellow2} \textbf{Yellow Weyl line cones.} (\textbf{\pana}-\textbf{\panc}) $E_\textrm{B}-k_{||}$ cuts sweeping perpendicular to the line node. In \pana\ we observe the upper cone associated with the double yellow line node; in \panb\ we see a line node crossing and the lower cone; in \panc\ we find a weak signature remaining from the lower cone. (\textbf{\pand}-\textbf{\pane}) MDCs taken from the $E_\textrm{B}-k_{||}$ cuts, as indicated by the green arrows in \pana-\panc. The weak peaks associated with the double yellow line node are marked by the red arrows. (\textbf{\pang}, \textbf{\panh}) The locations of the cuts in \pana-\panc, as marked. Our MDC analysis provides additional evidence for the yellow line node in Co$_2$MnGa.}
\end{figure*}

\begin{figure*}
\centering
\includegraphics[width=16cm,trim={1.1in 4.3in 1.1in 1in},clip]{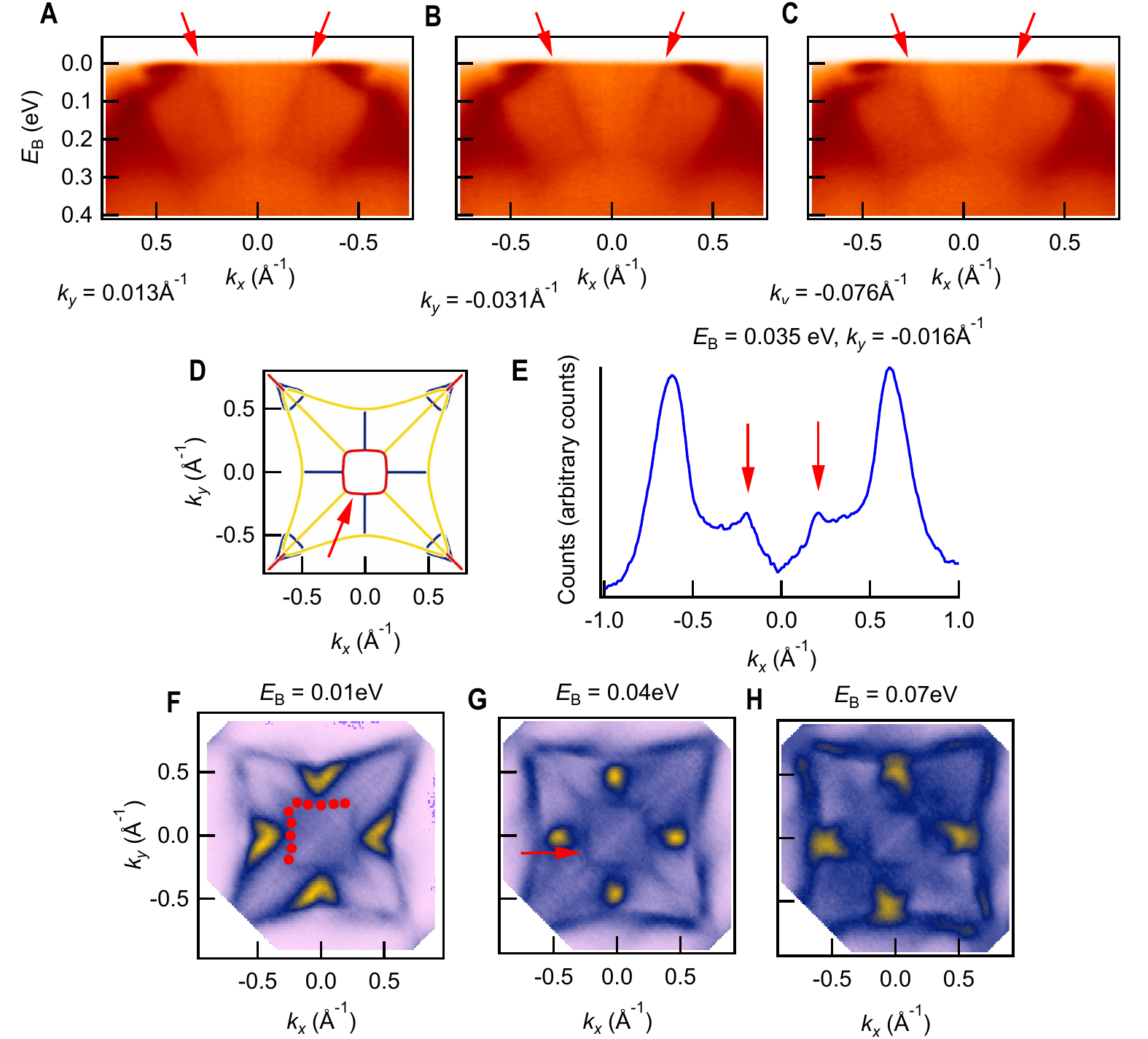}
\caption{\label{FigRed} \textbf{Signatures of the red Weyl line.} (\textbf{\pana}-\textbf{\panc}) $E_\textrm{B}-k_x$ cuts passing through the red line node near $\bar{\Gamma}$. We see two bands dispersing away from $\bar{\Gamma}$ as we approach $E_\textrm{F}$, consistent with (\textbf{\pand}) the red line node in calculation. (\textbf{\pane}) The two bands on an MDC (taken at the red arrow in {\pang}). (\textbf{\panf}-\textbf{\panh}) Constant-energy surfaces, emphasizing a square feature around $\bar{\Gamma}$ in agreement with the predicted red Weyl line. The square feature (red dots) appears to arise from the red Weyl line.}
\end{figure*}

\begin{figure*}
\centering
\includegraphics[width=16cm,trim={1in 5.4in 1in 1.35in},clip]{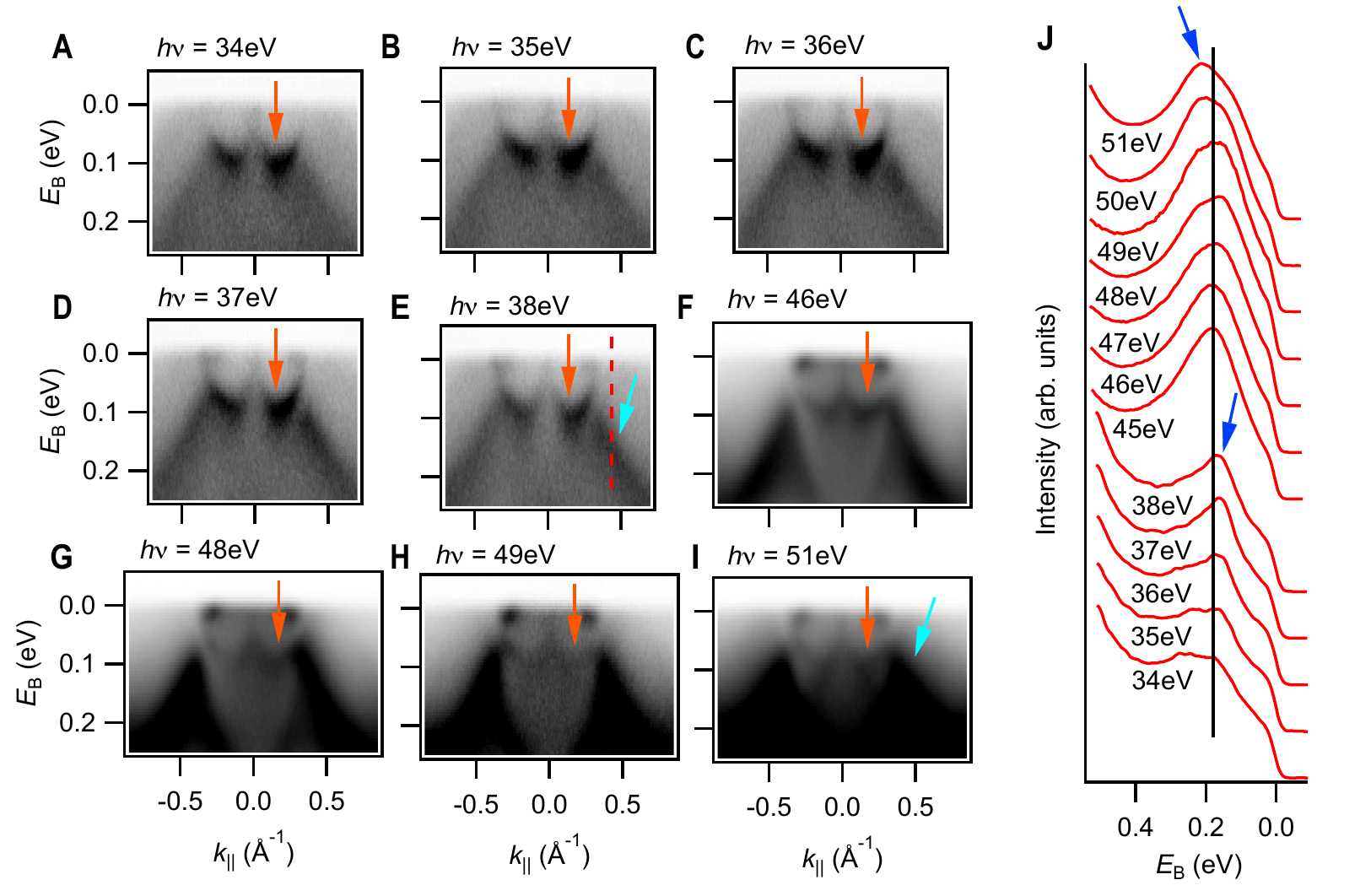}
\caption{\label{DH1} \textbf{Evidence for $k_z$ dispersion of the yellow Weyl line.} (\textbf{\pana-\pani}) $E_\textrm{B}-k_{||}$ cuts analagous to main text Fig. 3\pana-\panc, but at more photon energies. (\textbf{\panj}) Stack of EDCs as a function of $h\nu$, analogous to main text Fig. 3\pang, but instead of cutting through the drumhead surface state, the EDC cuts through the yellow line node, at $k_{||} = 0.45$ $\textrm{\AA}^{-1}$ (dotted red line in \pane). We clearly observe the drumhead surface state in all cuts (orange arrows). Recall that the drumhead showed no photon energy dependence, suggesting that it is a surface state. Here, by contrast, we see a photon energy dependence (blue arrows in \panj), associated with the yellow Weyl line (cyan arrows in \pane, \pani), suggesting a $k_z$ dispersion for the yellow line node cone.}
\end{figure*}

\begin{figure*}
\centering
\includegraphics[width=16cm,trim={1in 6.7in 1in 1.35in},clip]{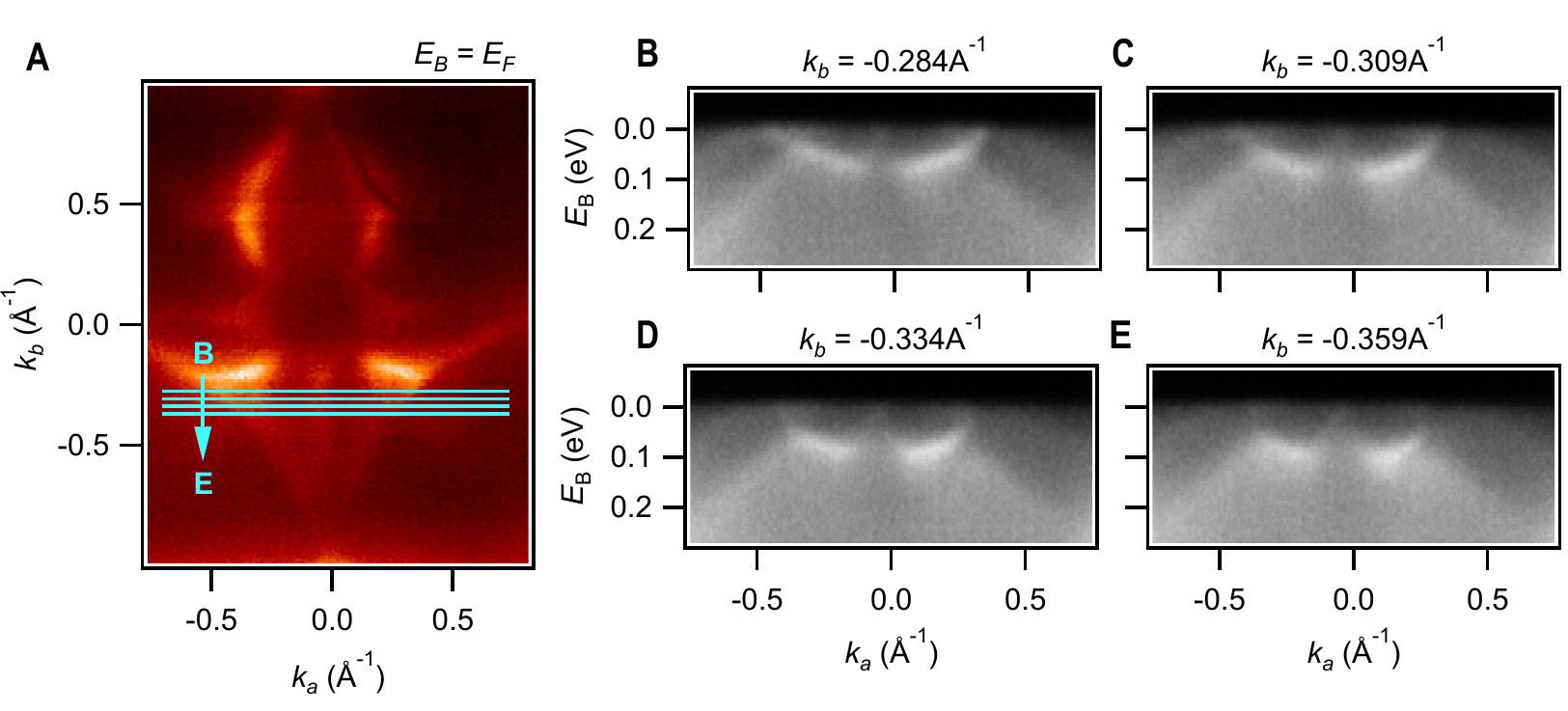}
\caption{\label{DH2} \textbf{In-plane dispersion of the drumhead.} (\textbf{\pana}) Fermi surface at $h\nu = 35$ eV and (\textbf{\panb-\pane}) $E_\textrm{B}-k_{||}$ cuts showing the drumhead surface state, with locations as marked in \pana\ (cyan lines). We observe a weak dispersion of the surface state doward in energy as we move away from $\bar{\Gamma}$. So we a observe an in-plane dispersion of the drumhead, but not an out-of-plane dispersion (main text Fig. 4), demonstrating a surface state.}
\end{figure*}

\begin{figure*}
\centering
\includegraphics[width=16cm,trim={1.1in 7.9in 1.1in 1.05in},clip]{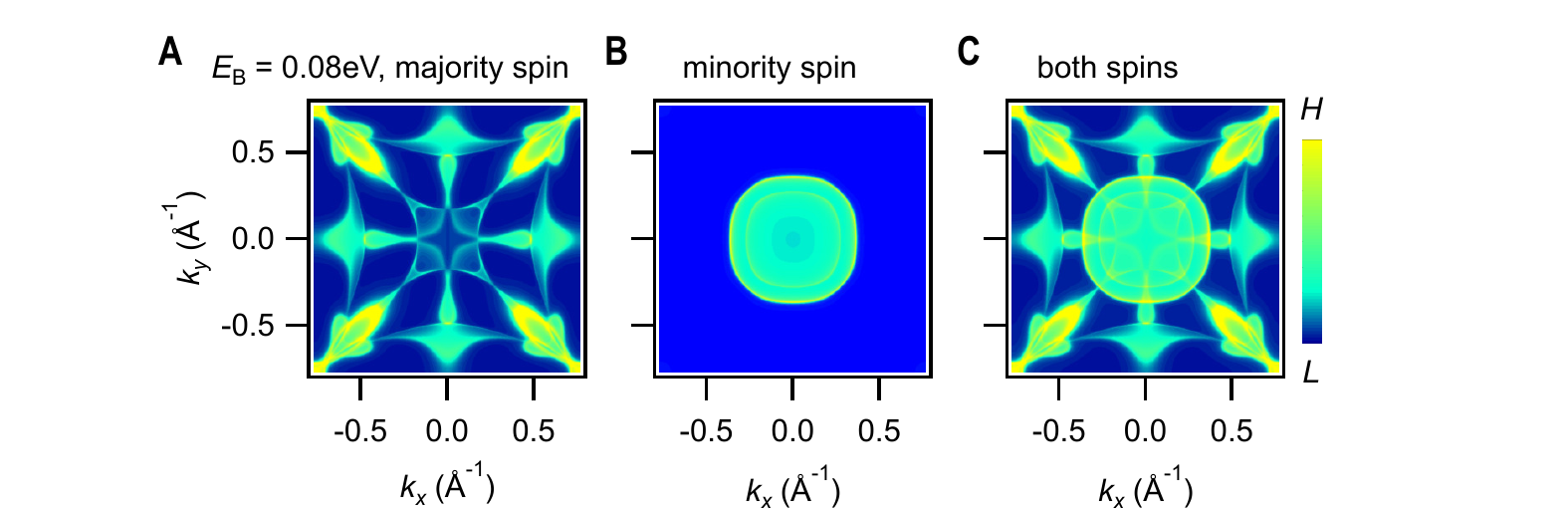}
\caption{\label{spinmin_calc} \textbf{Minority spin pocket in calculation.} ({\bf \pana}) Bulk projection of the majority spin states, same as main text Fig. 2\pane. ({\bf \panb}) Bulk projection of the minority spin at the same energy. ({\bf \panc}) The sum of \pana and \panb, the bulk projection of all states at the given binding energy.}
\end{figure*}

\begin{figure*}
\centering
\includegraphics[width=16cm,trim={1.1in 2.1in 1.1in 1.35in},clip]{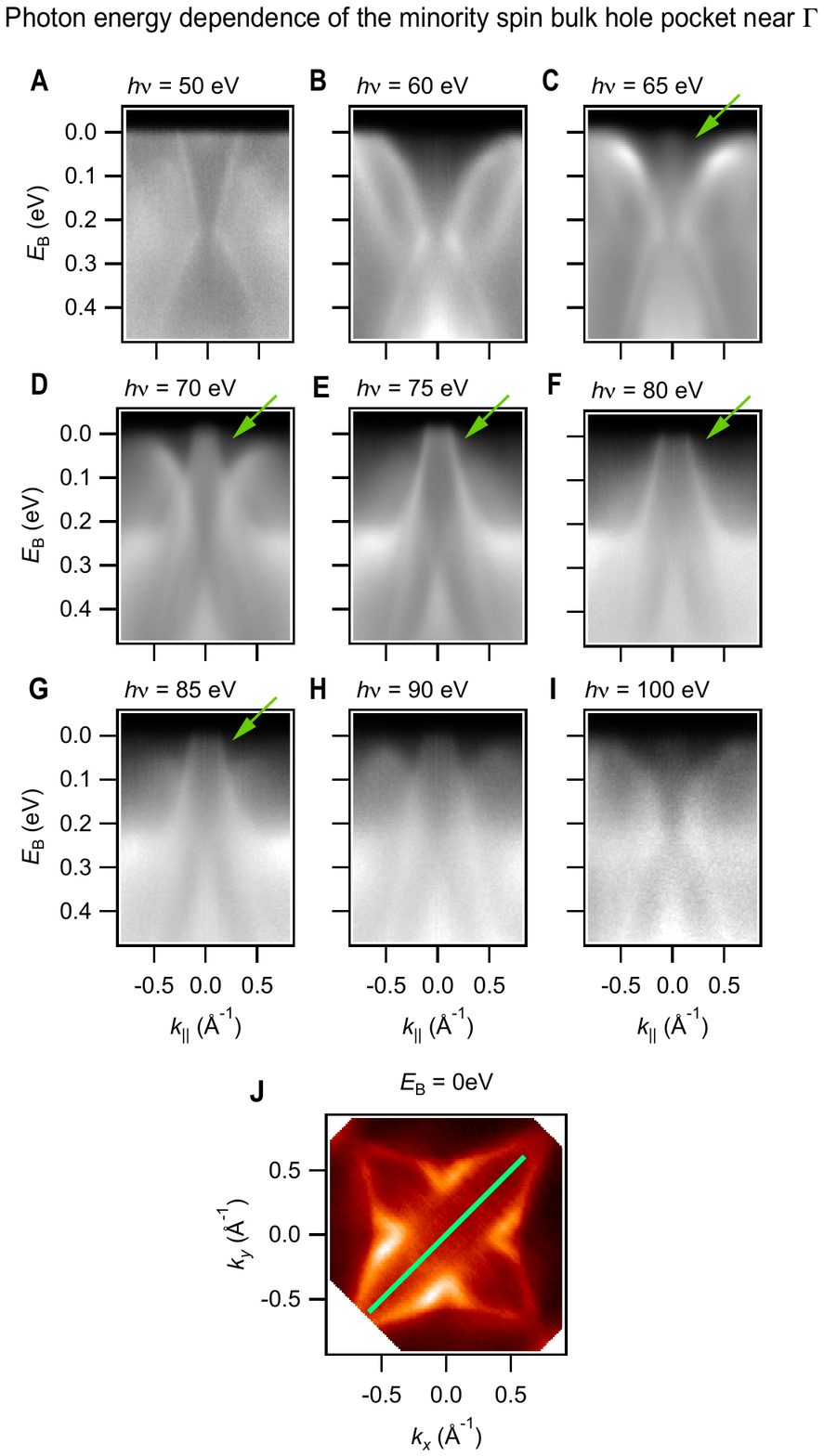}
\caption{\label{spinmin} \textbf{The minority spin pocket in ARPES.} (\textbf{\pana-\pani}) $E_\textrm{B}-k_a$ cuts through $\bar{\Gamma}$ at different photon energies, with the location of the cut shown in (\textbf{\panj}) by the green line. At  $h \nu > 65$ eV a large hole pocket appears (green arrows), consistent with the minority spin pocket seen in calculation.}
\end{figure*}

We briefly noted that Co$_2$MnGa has a large minority spin pocket around the $\Gamma$ point (Fig. \ref{FigSurvey}). We can omit this pocket from our ARPES measurements by an appropriate choice of photon energy $h\nu$, which then corresponds to a $k_z$ away from $\Gamma$. In particular, at $h \nu = 50$ eV, main text Fig. 2, we find that we cut near the top of the Brillouin zone (near the $X$ point) and far from the $\Gamma$ point. As a result, we then compare our data with the majority spin bands from \textit{ab initio}, as in main text Fig. 2\pane. However, to further compare ARPES and DFT, it is useful to search for this irrelevant pocket in ARPES and better understand why it does not compete with the line node and drumhead states in photoemission. In the DFT bulk projection, the minority spin projects onto a large pocket around $\bar{\Gamma}$, see Fig. \ref{spinmin_calc}\pana-\panc. Experimentally, we performed a photon energy dependence on a cut passing through $\bar{\Gamma}$. For $h \nu > 50$ eV we find that the red Weyl line disappears and a large hole pocket appears near $\bar{\Gamma}$ (Fig. \ref{spinmin}\panb-\panj). This pocket matches well with the minority spin pocket in calculation. This photon energy dependence suggests that the minority spin pocket does not interfere with our measurements of the Weyl lines because at $h\nu = 50$ eV we cut near the top of the bulk Brillouin zone in $k_z$ (near the $X$ point).

\markboth{}{}

\clearpage

\clearpage

\clearpage

\clearpage

\clearpage

\clearpage

\clearpage

\clearpage

\clearpage

\clearpage

\clearpage

\clearpage

\clearpage

\clearpage

\clearpage

\clearpage

\clearpage

\clearpage

\clearpage

\clearpage

\clearpage

\clearpage

\end{document}